\documentclass{aa}   

\usepackage{graphicx}
\usepackage{txfonts}
\usepackage{siunitx}

\AddToHook{begindocument/before}{\usepackage{hyperref}}

\def\Ha{\mbox{H$\alpha$}}
\def\Hb{\mbox{H$\beta$}}
\def\Msun{\mbox{${\rm M}_{\odot}$}}

\def\Rsun{\mbox{${\rm R}_{\odot}$}}

\def\kms{\mbox{km\,s$^{-1}$}}
\def\Teff{\mbox{$T_{\rm eff}$}}

\usepackage{xspace}
\usepackage{amsmath}

\newcommand{\spec}[3]{\mbox{#1\,{\sc #2}\,$\lambda$#3}\xspace}

\usepackage{xcolor}

\definecolor{beige}{HTML}{FFDF8C}

\begin{document}

   \title{Full orbital solution and dynamical masses
for a new shell-Be + sdOB binary AN~Col}

   \author{R.\ Klement
          \inst{1,2,3}
          \and
          Th.\ Rivinius
          \inst{1}
          \and
          A.\ C.\ Carciofi
          \inst{4}
          \and
          A.\ S.\ Chaturvedi 
          \inst{5}
          \and
          D.\ R.\ Gies 
          \inst{5}
          \and
          A.\ C.\ Fonseca Silva
          \inst{4}
          \and
          P.\ R.\ Querido
          \inst{4}
          }

   \institute{European Organisation for Astronomical Research in the Southern Hemisphere (ESO), Casilla 19001, Santiago 19, Chile \\
   \email{robertklement@gmail.com}
   \and
   The CHARA Array of Georgia State University, Mount Wilson Observatory, Mount Wilson, CA 91023, USA
   \and
   Université Côte d’Azur, Observatoire de la Côte d’Azur, CNRS, \\Boulevard de l’Observatoire, CS 34229, 06304 Nice Cedex 4, France
           \and
            Universidade de S\~ao Paulo, Instituto de Astronomia, Geof\'isica e Ci\^encias Atmosf\'ericas, \\ Rua do Mat\~ao 1226, Cidade Universit\'aria S\~ao Paulo-SP, 05508-090, Brazil
            \and Center for High Angular Resolution Astronomy, Department of Physics and Astronomy,\\ Georgia State University, P.O. Box 5060, Atlanta, GA 30302-5060, USA 
}

   \date{Received September 15, 1996; accepted March 16, 1997}

  \abstract
   {AN~Col is a shell Be star that has shown transient spectral features caused by a hot, evolved companion on a months-long orbit. If confirmed, this would place it among the growing category of post-mass-transfer spun-up Be + stripped subdwarf OB-type (sdOB) binaries. Characterizing the binary orbit enables obtaining dynamical masses, to provide a rare anchor for models of binary evolution and mass transfer.
   }
   {We present a full 3-dimensional orbital solution, dynamical masses, and dynamical parallax, for the newly discovered Be + sdOB binary AN~Col. As a prerequisite to obtain a reliable orbital solution, we need to characterize the disk-wide density oscillation cycle that gives rise to the dominant spectroscopic cycles in order to separate the weaker periodic orbital variations.
   }
   {We use near-IR interferometry time series from GRAVITY to confirm the companion and to map the relative astrometric orbit. With the knowledge of the orbital period, we identify spectral lines that trace the orbit of each of the two stellar components and we measure their radial velocities to obtain a double-lined spectroscopic orbit. Combining interferometry and spectroscopy, we derive the full 3-dimensional orbital solution.
   }
   {The disk has completed three full oscillation cycles since 2010, with a cycle length of $\sim 5.1$\,yr. The $\sim87.7$-day orbit is compatible with edge-on orientation, but the lack of eclipses in TESS light curves limits the orbital inclination to $88.4^{\circ}$ or less. The resulting masses are $M_{\rm Be} = 8.39\pm0.67$\,{\Msun} and $M_{\rm sdOB} = 0.81\pm0.22$\,\Msun. As in other Be + sdOB binaries, the estimated efficiency of the past mass transfer is unexpectedly high. Detection of a weak \spec{He}{ii}{4686} absorption and other \ion{He}{i} lines from the companion indicates that its effective temperature is close to $30$\,kK, and that it is a slow rotator.
   }
  {This work establishes AN~Col as a new benchmark Be + sdOB system, whose fundamental properties can be used to constrain models of binary evolution, as well as the physics of binary mass transfer that probably lead to the formation of the majority of Be stars. AN~Col is only the second Be + sdOB binary with a full orbital solution enabling the determination of parallax-independent dynamical masses for both the spun-up Be star and its stripped helium companion.
  }

   \keywords{Stars: massive -- stars: emission-line, Be -- circumstellar matter -- binaries: spectroscopic -- stars: individual: \object{AN\,Col}
               }

   \maketitle
   \nolinenumbers
%
\section{\label{sec:intro}Introduction}

Classical Be stars are rapidly rotating B-type stars which eject material in Keplerian disks as means to shed their excess angular momentum \citep[][and references therein]{2013A&ARv..21...69R, 2026enap....2..430R}. A prominent way of acquiring the excess angular momentum among Be stars, required to reach their near-critical rotation, is previous evolution in interacting binaries including a mass transfer episode from the originally more massive component onto the present-day Be star \citep{1991A&A...241..419P, 2024ARA&A..62...21M}. This evolutionary history can be confirmed by detecting the original mass donors, which for intermediate initial masses often appear as stripped subdwarf OB-type (sdOB) stars. 

\renewcommand{\arraystretch}{1.15}
\begin{table*}[t]
\caption{Spectroscopic datasets}
\label{tab:spectroscopy}
\centering
\begin{tabular}{c c c c c c c}
\hline\hline
Database & Instrument & Telescope & MJD range & $\lambda$ range [nm] & $R$ & Number\\
\hline
   - & CHIRON\tablefootmark{b} & SMARTS–GSU 1.5-m & 60638--61095 & 458--863 & 80000 & 19 \\
   ESO archive & UVES\tablefootmark{b} & VLT/UT2 8-m & 60219--60264 & 330--680 & $\sim40000$& 7 \\
   BeSOS & PUCHEROS & ESO 0.5-m  & 56736--57340 & 425--730 & $\sim17000$ & 7 \\ 
   BeSS & echelle\tablefootmark{a} & 0.30--0.36-m & 55853--60708 & 385--760 & 10000--11000 & 8 \\
   BeSS & non-echelle\tablefootmark{a} & 0.28-m & 57366--60187 & 650--662 & 12000--18000 & 5 \\
   ESO archive & XSHOOTER & VLT/UT3 8-m & 55467--56896 & 308--2400 & 4100--18400 & 6 \\
   - & TSPEC4\tablefootmark{b} & SOAR 4-m & 61100 & 9000--24000 & 3500 & 1 \\
   ESO archive & GRAVITY\tablefootmark{b} & VLTI/ATs 4$\times$1.8-m & 59877--60317 & 20000--24000 & 4000 & 8 \\ 
\hline
\end{tabular}
\tablefoot{
\tablefoottext{a} {Spectra taken by amateur astronomers with different small telescopes and spectrographs.}
\tablefoottext{b} {Original data.} 
}
\end{table*}

Stripped sdOB stars are much fainter than the bright Be stars, but their effective temperatures are typically higher, enabling their detection in the ultraviolet, and with more difficulty in optical spectra as well \citep{1981PASP...93..297P, 1998ApJ...493..440G, 2018ApJ...853..156W, 2023AJ....165..203W}. With sufficient contrast in the near-infrared---typically at least 200:1---and an angular separation of at least $\sim 0.5$\,milliarcsec (mas), sdOB companions to nearby Be stars become detectable with near-IR interferometry \citep{2015A&A...577A..51M, 2022ApJ...926..213K}. The combination of spectroscopic and astrometric orbits has thus far resulted in dynamical masses for about a dozen Be + sdOB binaries \citep{2022ApJ...940...86K, 2024ApJ...962...70K, 2025A&A...694A.172R, 2025A&A...694A.208K}. Because masses derived from evolutionary tracks depend on stellar evolution models, systems with model-independent dynamical masses provide the most stringent tests of binary evolution and mass-transfer calculations. Indeed, the new set of dynamical masses for Be + sdOB binaries has already resulted in new challenges to the expected values of mass transfer efficiency \citep{2025ApJ...990L..51L, 2025ApJ...987..210B}, and to the theory of stable mass transfer in general \citep{2026A&A...705A.225P}.

Even as the sample of well-characterized (binary) Be stars continues to steadily grow, detailed studies of individual systems remain essential due to the puzzling interplay of complex phenomena taking place. All Be stars display short-term photospheric variations from non-radial pulsations \citep{2003A&A...411..229R, 2016A&A...593A.106R, 2022AJ....163..226L}. These are connected to episodic mass ejections from the stellar photosphere that cause further short-term variations as they supply material to the circumstellar disk \citep{1998A&A...333..125R, 2025A&A...699A..82L}. The physical behavior of the disks largely confirms the predictions of the viscous decretion disk (VDD) model \citep{1991MNRAS.250..432L, 2006ApJ...639.1081C, 2008ApJ...684.1374C, 2011IAUS..272..325C}; similar to viscous accretions disks \citep{2019NewA...70....7M}, VDD attains nearly Keplerian rotation while transporting angular momentum outwards \citep[e.g.,][]{2018MNRAS.476.3555R}. The disks themselves are subject to a plethora of variations, such as cycles or episodes of large-scale growth and dissipation \citep{2012ApJ...756..156H, 2021ApJ...912...76M}, spiral wave density oscillations \citep{1991PASJ...43...75O}, and different variations caused by the (often otherwise invisible) orbiting companions \citep{2025A&A...698A.309R}.

Disk-wide density oscillations have been observed in about two thirds of Be stars, and the cyclic behavior typically proceeds on a time scale of years to decades \citep{1991PASJ...43...75O, 2009A&A...504..929S, 2009A&A...504..915C}. In a thin Keplerian disk, this oscillation takes the shape of a one-armed spiral with enhanced density, slowly precessing in prograde motion around the star \citep{1983PASJ...35..249K, 1991PASJ...43...75O, 1997A&A...318..548O, 2000ASPC..214..409O, 1992A&A...265L..45P, 2006A&A...456.1097P}. The main observable effect is the variation of the relative strength of the emission peaks in H$\alpha$ and other lines, as the denser part of the disk either approaches or recedes from the observer \citep{2013A&ARv..21...69R}. This can be quantified via the $V/R$ ratio of the maximum fluxes of the blue-shifted (violet) and red-shifted emission peaks relative to continuum.  Understanding and characterizing this long-term disk variability is essential since it can bias the radial velocity (RVs) offsets of spectral lines that are needed for a spectroscopic orbital solution.

As for the variations caused by companions, already the first confirmed Be + sdOB system --- another shell-Be star \object{$\phi$\,Per}, showed some hallmark spectral features that can serve as proxies for the presence and nature of the secondary \citep{1981PASP...93..297P}. Among those are orbital-phase-bound variations of the \ion{He}{i} lines, such as the occurrence of transient, sharp absorption/emission from specific \ion{He}{i} lines originating from metastable states. The characteristic orbital variations caused by a faint, hot companion are observed in AN~Col as well---these were initially discovered in archival BeSOS/PUCHEROS spectra, covering two distinct epochs: \spec{He}{i}{5015} and some other helium lines displayed sharp absorption components, but these were completely gone a little over a year later. On these grounds, we included AN\,Col in our interferometric observing program on detecting binary Be stars \citep{2024ApJ...962...70K}, as well as in our high-resolution spectroscopic monitoring program, leading to the dataset presented in this study.

\object{AN\,Col} (also HR\,1772, HD\,35165) has shown a spectrum compatible with a Be-shell stars for more than a century \citep{1953Obs....73...86B, 2006A&A...459..137R}. The spectrum was originally classified as about B5, but when using a spectrophotometric system that does not depend on spectral lines and thus avoid the complications caused by the shell nature, the data are more in agreement with an earlier type of B3\,IVe \citep{2018A&A...609A.108S}. Due to its relative faintness ($V=6.1$) compared to other nearby Be stars, like those in Orion and Canis Majoris, it is not a well observed or studied object. Up to now, no evidence for a close binary companion has been presented, although AN~Col was mentioned as a potential Be + sdOB binary by \citet{2022ApJ...926..213K} and \citet{2025A&A...694A.172R}. 

Among the roughly dozen Be + sdOB binaries with 3-dimensional orbital solutions, AN Col is only the second one to have a direct dynamical mass and distance determination not relying on a parallax measurement (after $\phi$~Per, \citealt{2015A&A...577A..51M}), and only the fourth one when also including the Be+pre-sdOB binaries HR~6819 \citep{2025A&A...694A.208K}, and V742~Cas \citep{2025ApJ...995..180G}. In its interplay of near edge-on orientation with companion-induced transient shell absorption features and a strong density oscillation cycle in its disk, AN~Col resembles previously studied Be + sdOB systems $\varphi$~Per \citep{2000A&A...358..208S, 2001A&A...368..471H}, FY~CMa \citep{2004A&A...427..307R}, HR~2142 \citep{2016ApJ...828...47P}, HD~55606 \citep{2018ApJ...865...76C}, and the first eclipsing Be binary V658~Car \citep{2025ApJ...995...87D}.

This work is structured as follows. The used observations are presented in Sects.~\ref{sec:obs}, \ref{appendix:tables}, and \ref{appendix:additional_observations}, the spectroscopic variability due to the $\sim5.1$\,yr disk density oscillation cycle is characterized in Sects.~\ref{sec:spectroscopy} and \ref{appendix:vr_cycle}, and the companion-induced spectroscopic variability is described in Sects.~\ref{sec:companion_induced_variability} and \ref{appendix:additional_transient_shell}. The orbital solution combining interferometry and spectroscopy is presented in Sect.~\ref{sec:orbit}, and the results are discussed Sects.~\ref{sec:discussion} before the main conclusions are given in Sect.\ref{sec:concl}.

\section{\label{sec:obs}Observations}

\subsection{\label{sec:obs:spec}Spectroscopy}

We collected 26 moderate-resolution spectra from amateur and public archives (BeSOS, BeSS, and ESO archive for XSHOOTER), as well as 26 new high-resolution spectra from our own observations (CHIRON, UVES), with the entire dataset spanning 15.6 years. This is complemented by one near-IR spectrum from TripleSpec4.1 (TSPEC4), and eight near-IR $K$-band spectra extracted from GRAVITY interferometry (Sect.\ref{sec:obs:intf}).
See Table~\ref{tab:spectroscopy} for a summary of all used spectra.

The CTIO HIgh ResolutiON \citep[CHIRON;][]{tokovinin_chironfiber_2013} echelle spectrograph is mounted on the 1.5m SMARTS-GSU telescope at the Cerro Tololo Inter American Observatory (CTIO). All of our spectra were obtained using the "slicer" mode which splits the light across 59 echelle orders with a resolving power of $R\approx80\,000$. The spectra are delivered as flat-fielded, bias-subtracted, and wavelength-calibrated. These spectra were then corrected for the instrumental blaze function by fitting a 5th-order polynomial to each echelle order and dividing out the continuum. The deblazed orders were then merged by taking a weighted average of overlapping contributions and interpolating across gaps between orders. The final merged spectra were interpolated onto a common heliocentric uniform log wavelength grid with a pixel size of 5~km s$^{-1}$ pixel$^{-1}$ using \texttt{IDL}. As the CHIRON data are somewhat noisy at this sampling, we filtered the spectra with a Gaussian of ${\rm FWHM} = 0.3$\,{\AA} for measurements and display.

UVES is a high-resolution echelle spectrograph, covering the near-UV to the near-IR domain \citep{2000SPIE.4008..534D}, though not doing so in a single setup but via a number of cross-disperser settings. The central wavelength of the cross-disperser setting is used to identify the instrument mode. AN\,Col was observed in the 390/580 setting, giving data from 330 to 450\,nm in the blue arm and 480 to 680\,nm in the red arm, the latter having a gap between 575 and 585\,nm. All spectra were observed with 1{\arcsec} slit width, delivering a resolving power of $R\equiv\Delta\lambda/\lambda\approx40\,000$. The UVES data were reduced with the ESO pipeline workflows provided under ESOReflex \citep{2013A&A...559A..96F}.

Be Stars Observation Survey\footnote{\url{https://besos.ifa.uv.cl/}} \citep[BeSOS;][]{2018MNRAS.474.5287A} is a database of spectra taken with the PUCHEROS spectrograph \citep{2012MNRAS.424.2770V}. The BeSS database\footnote{\url{http://basebe.obspm.fr/basebe/}} collects amateur observations, and is thus inhomogeneous in the instruments and telescopes used, covered range, resolution, and quality. 

XSHOOTER \citep{2011A&A...536A.105V} is an echelle spectrograph with three independent arms covering a wide wavelength range from 300 to 2480\,nm in medium spectral resolution. First epoch slit width was 1.3\arcsec in UVB, and 1.2\arcsec in VIS and NIR, corresponding to spectral resolving power of 4100, 6500, and 4300 for the three bands, respectively. The second epoch used 0.5{\arcsec} in UVB, and 0.4{\arcsec} in VIS and NIR, corresponding to spectral resolving power of 9700, 18400, and 11600. The data were reduced with the ESO pipeline workflows provided under ESOReflex \citep{2013A&A...559A..96F}.

TSPEC4 \citep{2014SPIE.9147E..2HS} is a near-IR (0.94–2.46 $\mu$m) cross-dispersed spectrograph on the 4.1-m Southern Astrophysical Research (SOAR) telescope. Our observing strategy consisted of two ABBA nodding sequences, using individual exposure times of 5 seconds per slit position (totaling 40 seconds of integration time). 

\begin{figure}[t]
\begin{center}
\includegraphics[width=\hsize]{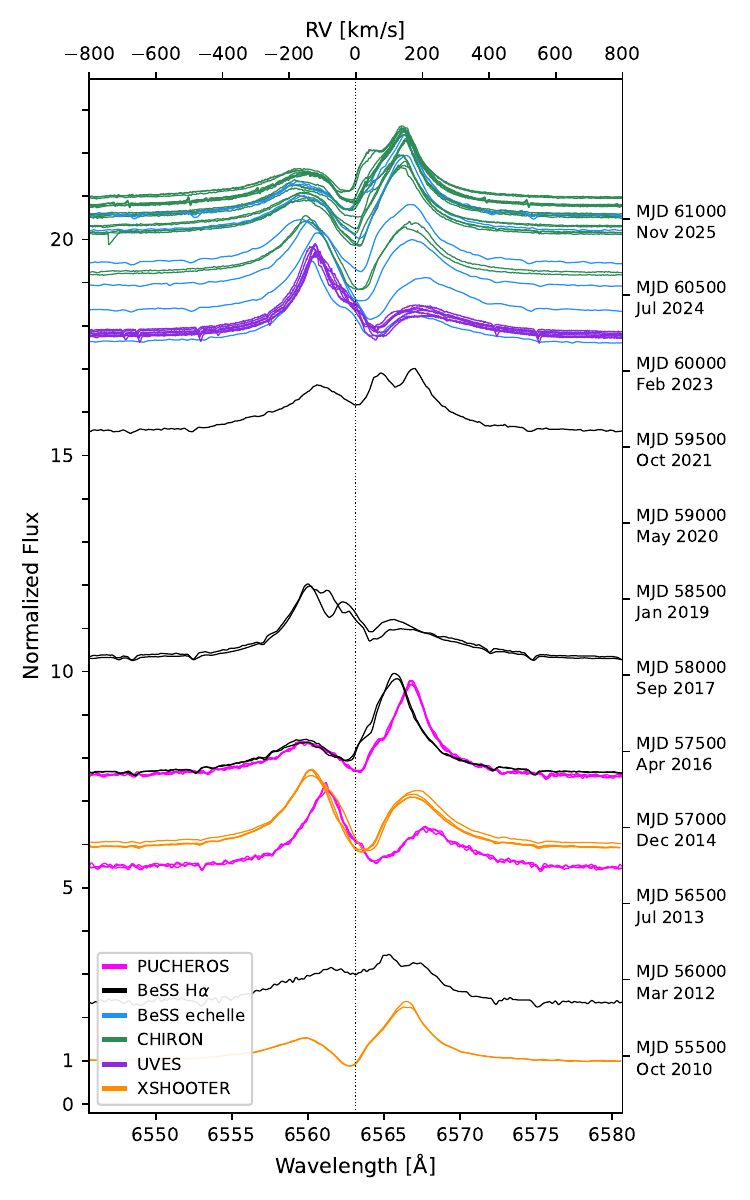}%
\end{center}
\caption[xx]{\label{fig:spectra_halpha_all}Normalized H$\alpha$ line profiles covering the last $15.6$ years. Roughly three full $V/R$ disk oscillation cycles are covered.}
\end{figure}

\begin{figure}[ht!]
\begin{center}
\includegraphics[width=\hsize]{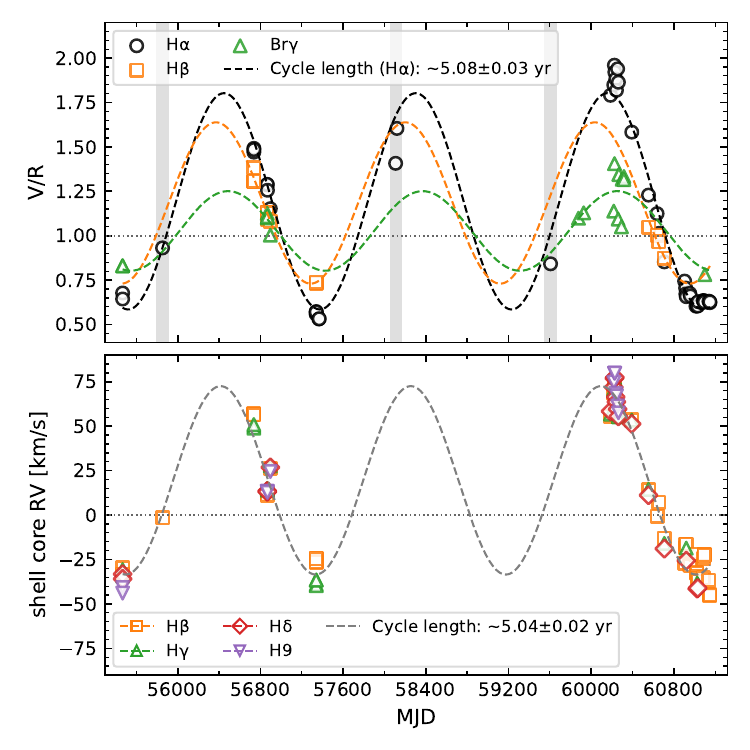}%
\end{center}
\caption[xx]{\label{fig:EW_RV_VR_stacked} \textit{Upper panel:} The measured $V/R$ ratios for {\Ha}, {\Hb}, and Br$\gamma$ and corresponding sine fits. Epochs when triple-peaked H$\alpha$ was observed are highlighted by a gray span.
\textit{Lower panel:} Shell core absorption RVs for {\Hb}, H$\gamma$, H$\delta$, and H9 with a combined sine fit, corrected for the systemic velocity. 
}
\end{figure}

\subsection{\label{sec:obs:intf}Interferometry}

A total of eight snapshots were collected with the GRAVITY 4-telescope beam combiner operating in the near-IR $K$ band at the Very Large Telescope Interferometer (VLTI). VLTI was used with the Auxiliary Telescope (AT) Array in the `Large' and `Extended' configurations. The maximum baseline between the telescopes ($B_{\rm max})$ for these configurations correspond to $K$-band angular resolution $\lambda/2B_{\rm max}$ of $\sim1.7$, and $\sim1.1$ milliarcsec (mas), respectively. The data were collected under the most relaxed atmospheric conditions constraints: seeing $1.4$\,arcsec or better, and coherence time in the visible of 1.6\,ms or more. 

GRAVITY was used in the single-field on-axis mode, in which the light from the science target is split between the science beam combiner (SC) and the fringe tracker (FT). We used an Object -- Sky -- Object observation sequence with an integration time of 100\,s in the SC channel and a total of four exposures for each Object/Sky observation. The spectral resolution was set to High ($R=4000$) and the polarization mode to Combined. Each observation of the science target was followed by observation of the calibrator star HD\,35554 with a uniform diameter $0.1105\pm0.0042$\,mas \citep{2017yCat.2346....0B} to enable absolute calibration of the atmospheric transfer function. The total observation time including overheads per calibrated snapshot was one hour. The data were reduced and calibrated with the official ESO GRAVITY data reduction pipeline version 1.7.0 in the ESOReflex environment  \citep{2013A&A...559A..96F}. Dates and telescope baselines for our observations are listed in Table~\ref{tab:astrometry} in Appendix~\ref{appendix:tables}.

Each calibrated observation consists of a set of absolute visibilities (|V|), closure phases (T3PHI), as well as differential phases (DPHI), and a normalized $K$ band spectrum (NFLUX). After careful visual inspection of the data revealed some inconsistencies between the two Object scans for the 2023~Nov~12,  2023~Dec~14, and 2023~Dec~21 epochs, we discarded the inferior Object scan (noisier, visibly miscalibrated |V|) for each of these snapshots. The calibration inconsistencies were almost certainly caused by variable seeing conditions, as these data were collected under challenging atmospheric conditions.

\subsection{Optical polarimetry}
\label{sec:obs:polarimetry}

Broad-band linear polarimetric observations of AN\,Col were obtained on
2024 September 24 and October 29 with the IAGPOL imaging polarimeter
\citep{Magalhaes1996} at the Observatório do Pico dos Dias (OPD/LNA,
Brazil). IAGPOL employs a rotating half-wave plate followed by a calcite
analyzer, allowing simultaneous measurements of the ordinary and
extraordinary beams. Observations were acquired in the Johnson--Cousins
$BVRI$ filters.
The data were reduced using the standard IAGPOL pipeline
\citep{Magalhaes1996}.

\subsection{Other observations}\label{sec:obs:otherobs}

AN\,Col (WDS J05213-3421A) has a third, wide companion (component B) which is about 4 magnitudes fainter than the inner Be + sdOB binary \citep{2019yCat....102026M}. The full set of six observations that resolved the wide binary---taken between 1922 and 1969---resulted in a range of $1.78$ to $2.24"$ for the angular separation, and $5.6$ to $8.8^{\circ}$ for the position angle, with no apparent orbital movement. In Appendix~\ref{appendix:gaia}, we analyze the recent Gaia DR3 data, according to which the inner binary and the wide companion are gravitationally bound, but the derived Gaia distances do not agree perfectly with the orbital parallax for the inner Be + sdOB pair derived in this study. In Appendix~\ref{appendix:astrometry_from_acquisition_cameras}, we use data from acquisition cameras of different instruments in order to assess possible changes in the relative astrometry of the B component in recent decades, but we fail to detect any significant orbital motion.

We also analyzed photometry from TESS, with data available from three separate sectors (Appendix~\ref{sec:TESS}). Considering the near edge-on orbit, photometric eclipses could not be identified in TESS data, although phase-bound, but transient variations may be present (Fig.~\ref{fig:TESS}).

\section{\label{sec:spectroscopy}Long-term $V/R$ cycle}

\subsection{Spectroscopic analysis}

The observed spectra of AN~Col span almost 16 years (2010~Sep to 2026~Apr) and show complex variations on two main time scales. The stronger, $\sim5.1$-yr cyclic variations come from large-scale density oscillation in the Be disk, while the less conspicuous $87.7$-day periodic variations arise from the influence of a stripped sdOB companion which is hotter than the Be star (Sect.~\ref{sec:companion_induced_variability}). In order to arrive at a reliable orbital solution, it is a prerequisite to first characterize the large cyclic variations from the $V/R$ cycle.

\subsubsection{H$\alpha$ and other hydrogen lines}

\begin{figure*}[th]
\centering

\begin{minipage}{0.195\textwidth}
    \centering
    \includegraphics[width=\linewidth]{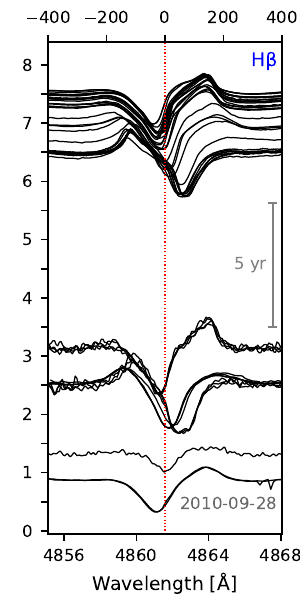}
\end{minipage}
\hfill
\begin{minipage}{0.195\textwidth}
    \centering
    \includegraphics[width=\linewidth]{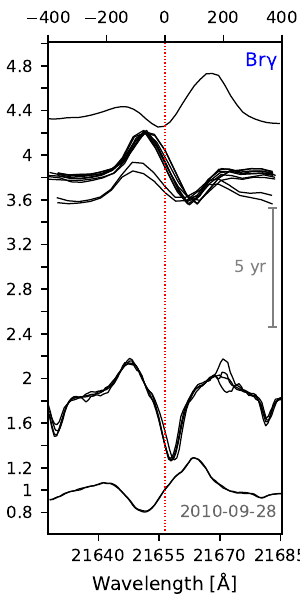}
\end{minipage}
\hfill
\begin{minipage}{0.195\textwidth}
    \centering
    \includegraphics[width=\linewidth]{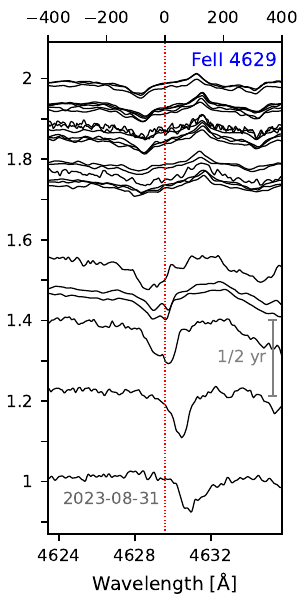}
\end{minipage}
\hfill
\begin{minipage}{0.195\textwidth}
    \centering
    \includegraphics[width=\linewidth]{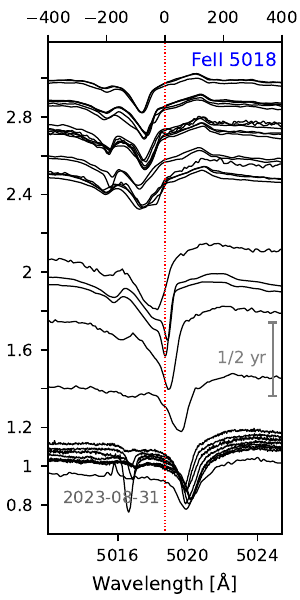}
\end{minipage}
\hfill
\begin{minipage}{0.195\textwidth}
    \centering
    \includegraphics[width=\linewidth]{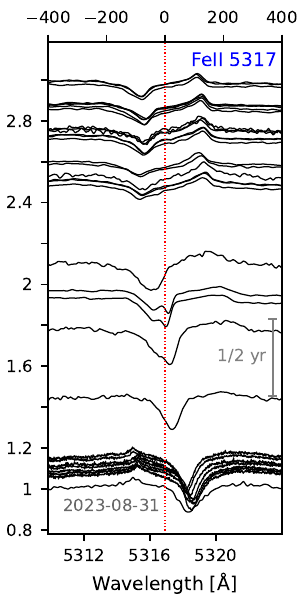}
\end{minipage}

\caption{\label{fig:spectra_hydrogen_all} Selected time series from the spectroscopic dataset. In each subplot, the vertical offsets between spectra are proportional to the epochs of the individual observations, with the vertical line showing the absolute time scale, and the printed date giving the epoch of the first spectrum in the time series. The upper $x$ axis shows RVs in \kms, corrected for the systemic velocity.
\textit{Left two panels:} The entire time series of H$\beta$ and Br$\gamma$ spectra.
\textit{Right three panels:} Profiles of selected \ion{Fe}{ii} lines from the UVES part of our spectroscopic dataset onwards.
}
\end{figure*}

In our dataset, H$\alpha$ is observed to go through three complete cycles of $V<R \rightarrow$ Triple peak $\rightarrow V>R \rightarrow V\approx R$, with the latest spectra capturing a $V<R$ phase again, and the two very last ones showing a possible onset of the next triple-peaked phase (see Fig.~\ref{fig:spectra_halpha_all}). The prominence of the specific $V$ or $R$ emission peaks in H$\alpha$ is accompanied by a respective shift of the central absorption redwards (when $V>R$) or bluewards ($V<R$). The $V/R$ ratio, measured by fitting narrow Gaussians to the $V$ and $R$ emission peaks (ignoring any additional central peak, see below), has varied between $\sim0.5$ and $\sim1.9$, and a simple sine wave fit indicates a cycle length of $5.08\pm0.03$\,yr ($1855\pm10$\,d, Fig.~\ref{fig:EW_RV_VR_stacked}, top panel).

The $V$ and $R$ peaks reach above continuum also in the near-IR Br$\gamma$ line (Fig.~\ref{fig:spectra_hydrogen_all}, second panel) and other lines in the Brackett and Paschen series, and sometimes in H$\beta$. While the dataset is scarcer for these lines, the $V/R$ phase of H$\beta$ seems to precede that of H$\alpha$ by about 5.4\% of the cycle length, and Br$\gamma$ trails it by about 3.6\%. An opposite trend in regards to Br$\gamma$ was observed in $\zeta$~Tau, where the Br$\gamma$ $V/R$ phase precedes that of H$\alpha$ rather than trails it \citep{2009A&A...504..929S, 2009A&A...504..915C}. Dynamical plots of {\Ha} and {\Hb} lines phased with the cycle length are shown in Fig.~\ref{fig:dynspec_vr_phased}.

Three spectra in our dataset show a clearly defined triple-peaked emission in H$\alpha$, with one more showing a transitory profile right after the second observed triple-peaked phase (Fig.~\ref{fig:triple_peak}). The triple-peaked phases are observed to occur only when the $V/R$ ratio is increasing, i.e. during transitions from $V<R$ to $V>R$, but not during the reverse transitions \citep[the same was observed for $\zeta$~Tau;][see Sect.~\ref{sec:discussion} for a detailed discussion and comparison to other stars]{2009A&A...504..929S}. The transitory spectrum taken on MJD $58122$---in which the middle peak observed shortly before (MJD $58109$) has apparently split into two smaller ones before disappearing---precedes the maximum of the $V>R$ phase. The notable difference in these two spectra taken shortly apart suggests a more complex variability during this phase than captured by the otherwise mostly smooth $V/R$ curve. The two other triple-peaked spectra (MJD $55854$ and $59607$) occurred at a similar $V/R$ phase two cycles apart, and indeed show similar profiles (Fig.~\ref{fig:triple_peak}). 

The RVs of the central shell absorption can be better measured from H$\beta$ and higher Balmer lines rather than H$\alpha$, where the central absorption is shallow and wide. The shell core RVs for H$\beta$, H$\gamma$, H$\delta$, and H9 hydrogen lines---selected as representative examples---have varied between $-30$ and $+100$\,{\kms} with the same cycle length as that of the $V/R$ cycle (Fig.~\ref{fig:EW_RV_VR_stacked}, bottom panel showing combined sine fit). When fitted separately, the three higher lines precede H$\beta$ in the phase of the shell core RV cycle by about 1\%, 2\%, and 6\% of the cycle length, respectively.

\subsubsection{Metal lines}

Other spectral lines display shell absorption with similar long-term variation as hydrogen lines, as well as small-scale emission peaks coinciding with the dominant emission peaks in \Ha. Among the lines exhibiting this are multiple \mbox{Fe\,{\sc ii}} lines such as for instance \spec{Fe}{ii}{4629}, \spec{Fe}{ii}{5018}, and \spec{Fe}{ii}{5317}  (Fig.~\ref{fig:spectra_hydrogen_all}, right three panels). An emission peak is clearly seen at certain epochs in \spec{Fe}{ii}{5317}, \spec{Fe}{ii}{6319}, and others. There are also several \mbox{Fe\,{\sc ii}} emission-only fluorescent features in the near IR that show clear and often double-peaked emission, among these \spec{Fe}{ii}{9997} and \spec{Fe}{ii}{10500} \citep[very similar features were seen in the near-IR spectra of the eclipsing Be + sdOB binary V658 Car;][]{2025ApJ...995...87D}. Prominent lines of other elements showing long-term variations include \spec{Ca}{ii}{3934}, \spec{Ca}{ii}{3968}, \spec{Mg}{ii}{4481}, \spec{Na}{i}{5890}, \spec{Na}{i}{5896}, the \spec{O}{i}{7774} triplet, and the Ly$\beta$-pumped \spec{O}{i}{8446}. The last mentioned line shows a double-peaked emission similar to the near-IR fluorescent lines of \mbox{Fe\,{\sc ii}}. The infrared \ion{Ca}{ii} triplet is not present in emission.

The first two epochs of CHIRON spectra (MJD~$61036$ and $60960$) capture the transition in the $V/R$ cycle from $V>R$ to $V<R$, and from the central shell absorption being redshifted to being blueshifted, which is well seen in \spec{Fe}{ii}{4629}, \spec{Fe}{ii}{5018}, and \spec{Fe}{ii}{5317} (Fig.~\ref{fig:spectra_hydrogen_all}, right three panels). This helps with revealing a bimodal behavior of the shell absorption -- the deeper absorption occurs when redshifted, and diminishes when its RV offset decreases to $\sim+25$\,{\kms}. Just before diminishing, but while still prominent, a shallower absorption appears blueshifted at $\sim-25$\,{\kms}. In summary, the shell absorption is sharper and deeper when redshifted, and does not occur between $\sim-25$ and $\sim+25$\,{\kms}.

\subsection{Spectro-interferometric analysis}

The Be star disk was angularly resolved in the Br$\gamma$ emission, as seen in our high-spectral-resolution ($R=4000$) GRAVITY data. The first two GRAVITY snapshots were taken some time after the last observed triple-peaked {\Ha} phase as $V/R$ was still rising, while the other six were taken just after the $V>R$ maximum, around the same time as the UVES spectra. The Br$\gamma$ line profiles from GRAVITY data are displayed in the upper part of the second panel of Fig.~\ref{fig:spectra_hydrogen_all} (except for the uppermost spectrum which is from TSPEC4).

The differential phases in the Br$\gamma$ line display slightly asymmetric S-shaped profiles, confirming that the Be disk displays an actual geometrical asymmetry on top of Keplerian rotation. The spectro-interferometric Br$\gamma$ data show asymmetric signatures also in the absolute visibilities, and less prominently in the closure phases, reflecting the disk oscillation stage at the time of the observations. These asymmetries present a challenge when fitting for the Keplerian disk parameters, as introducing a dedicated disk oscillation model for AN~Col like was done for $\zeta$~Tau by \citet{2009A&A...504..915C} is beyond the scope of this paper. For the analysis below, we selected the first epoch of GRAVITY data, which shows the most symmetric line profile (Fig.~\ref{fig:spectra_hydrogen_all}).

Despite the asymmetry, the data clearly indicate close to edge-on orientation of the disk major axis (as must be the case for Be-shell stars), and a disk position angle of $\sim 275^{\circ}$ (measured from north to east and corresponding to the receding side of the disk). As expected, these results correspond to the same geometry as that of the astrometric orbit (Sect.~\ref{sec:astrometry_fitting}). The FWHM of the Br$\gamma$ line-emitting region is around $1.5$\,mas, i.e., about a half of the orbital semimajor axis, although the disk itself extends further and possibly even beyond the binary orbit \citep{2017A&A...601A..74K, 2019ApJ...885..147K, 2025A&A...698A.309R}.

The width of the Br$\gamma$ line and the corresponding spectro-interferometric signature also require a rotational velocity at the base of the Keplerian disk of at least $\sim 500$\,{\kms} which is higher than what can be derived from the literature $v_{\rm rot} \sin{i}$ of $240\pm5$\,{\kms} \citep{2018MNRAS.474.5287A} for close to edge-on orientation. However, in our experience, $v_{\rm rot} \sin{i}$ values for shell stars tend to be underestimated in the literature\footnote{This conclusion is based on spectra of several Be-shell stars obtained during episodes of weak or absent disks over the past few decades.}. Comparing the data to synthetically broadened spectra of \spec{He}{i}{6678} at $\Teff=25$\,kK and $\log g=4$, the width of the broad photospheric Be star component---when ignoring the shell contribution---agrees significantly better with values between 350 and 400\,{\kms} (detailed determination is not attempted here). Considering the ratio $W=v_\mathrm{rot}/v_\mathrm{orb}$, where $v_{\rm orb}$ is the Keplerian speed at the equator, our $v_{\rm rot} \sin{i}$ estimate and the observed $v_{\rm orb}$ result in $W$ between 0.7 and 0.8, which are the typical values for Be stars at large \citep{2013A&ARv..21...69R}. With $W=0.8$, these numbers are consistent with the candidate rotational frequency of $\sim1.84$\,c/d (Sect.~\ref{sec:TESS}) for a polar radius of $\sim3.3$\,{\Rsun} (equatorial radius of $\sim4.3$\,{\Rsun}), which is somewhat lower than polar radius expected for a B2-type MS star \citep[e.g. $\sim4.1$\,{\Rsun} according to ][]{2013ApJS..208....9P}.

An asymmetric phase profile from a spiral wave-like density pattern can be used to determine whether we see the disk from above (rotating counter-clockwise) or from below (rotating clockwise). This was successfully done for $\zeta$~Tau \citep{2009A&A...504..915C}, and can in principle be done for AN~Col as well, but we currently lack a dedicated oscillation model. Nevertheless, if the disk is aligned with the orbit (Sect.~\ref{sec:astrometry_fitting}), we should be seeing the disk from above, rotating counter-clockwise.

\subsection{Polarimetric analysis}

Polarimetric observations provide independent confirmation of the disk position angle inferred from the GRAVITY interferometry. Table~\ref{tab:pol} lists the observed broad-band linear polarization in the $BVRI$ filters. Although no direct estimate of the interstellar polarization (ISP) is available for AN\,Col, it can be constrained from the nearby star HD\,34677. \citet{Heiles2000} measured a $V$-band polarization of $0.040\pm0.044$\% for this star, which lies only $1.17^\circ$ from AN\,Col and at a comparable distance. We therefore assume that the ISP toward AN\,Col is negligible and that the observed polarization is dominated by the intrinsic component. 

The polarization degree exhibits a wavelength dependence characteristic of Be-star disks \citep{Haubois2014}, along with modest variability between the two observing epochs. In contrast, the polarization position angle remains remarkably stable. Averaging all eight measurements listed in Table~\ref{tab:pol} yields position angle of $5.4 \pm 2.5^\circ$, where the uncertainty is the standard deviation of the measurements. In Be-star disks, linear polarization produced by electron scattering is expected to be oriented perpendicular to the disk plane \citep{Wood1996,Halonen2013}. The corresponding disk position angle inferred from the polarimetry is therefore $95.4^\circ$, or equivalently $275.4^\circ$ under the $180^\circ$ ambiguity, in excellent agreement with the disk position angle of $\sim275^\circ$ determined from the GRAVITY spectro-interferometric observations.

\section{Companion-induced spectroscopic variability}\label{sec:companion_induced_variability}

\begin{figure*}[th]
\centering

\begin{minipage}{0.195\textwidth}
    \centering
    \includegraphics[width=\linewidth]{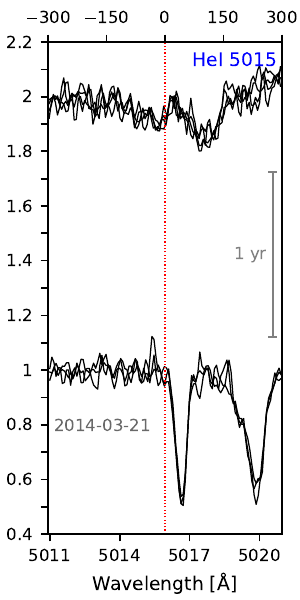}
\end{minipage}
\hfill
\begin{minipage}{0.195\textwidth}
    \centering
    \includegraphics[width=\linewidth]{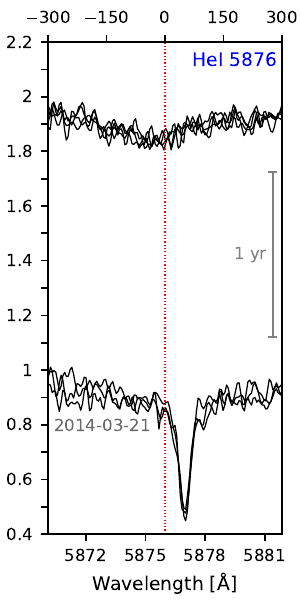}
\end{minipage}
\hfill
\begin{minipage}{0.195\textwidth}
    \centering
    \includegraphics[width=\linewidth]{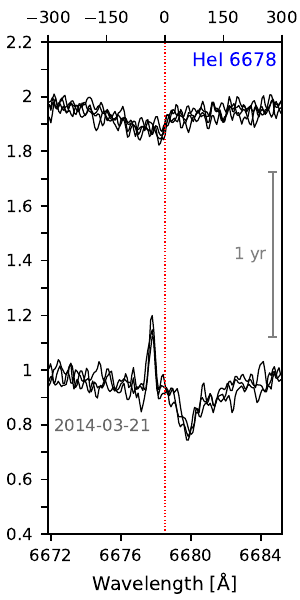}
\end{minipage}
\hfill
\begin{minipage}{0.195\textwidth}
    \centering
    \includegraphics[width=\linewidth]{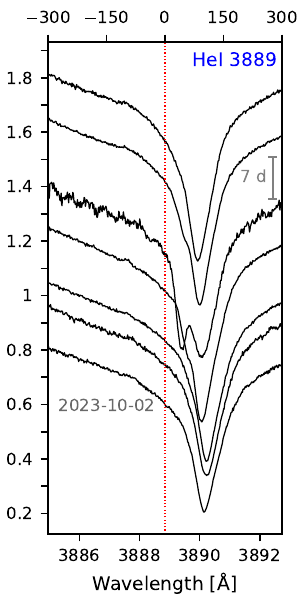}
\end{minipage}
\hfill
\begin{minipage}{0.195\textwidth}
    \centering
    \includegraphics[width=\linewidth]{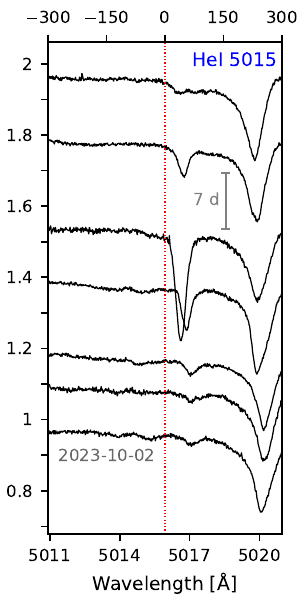}
\end{minipage}

\caption{\label{fig:transient_shell_BESOS_UVES} Same as Fig.~\ref{fig:spectra_hydrogen_all}, but showing the transient occurrence of the companion-induced shell absorption (and emission) as seen in the BESOS/PUCHEROS (\textit{left three panels}, transient shell at orbital phase $\phi=0.94$) and UVES spectra (\textit{right two panels}, transient shell visible between $\phi=0.85$ and $\phi=1.05$). For further illustration of the transient shell, see Fig.~\ref{fig:transient_shell_overplot}.
}
\end{figure*}

Several epochs capture the occurrence of transient shell absorption in \ion{He}{i} lines, most prominently in \spec{He}{i}{5015} and \spec{He}{i}{5876}, and to a lesser extent in \spec{He}{i}{4471}, \spec{He}{i}{6678}, and \spec{He}{i}{7281}. The transient shell absorption is interpreted as a line-of-sight effect caused by the sdOB companion heating up the outer parts of the Be disk that are projected onto the Be star photosphere. Similar transient shell phases caused by the same type of companion have been observed in several other Be stars (see Sect.~\ref{sec:discussion} for a discussion and comparison).

The transient shell was always observed close to the inferior conjunction as the sdOB passed in front of the Be star (at orbital phases after 0.82 and before 0.08 using the solution presented in Sect.~\ref{sec:orbit}, see Fig.~\ref{fig:orbit_transient_shell}). The RV of the shell appears to remain relatively stable during a single orbit, while in the long term its absolute RV offset and strength follows that of the main shell absorption as it varies during the $V/R$ cycle---from stronger red-shifted absoroption while $V>R$ to weaker blue-shifted absorption while $V<R$. The appearance of the transient shell in \spec{He}{i}{5015} in the two distinct $V/R$ phases is shown in Fig.~\ref{fig:transient_shell_overplot}. 

The first prominent transient shell occurs at the first PUCHEROS epoch (MJD $56737$, orbital phase $\phi=0.94$), which displays a sharp absorption at $\sim+40$\,{\kms} (corrected for systemic velocity, see Fig.~\ref{fig:transient_shell_BESOS_UVES}, left three panels). Another transient shell absorption phase occurred during our time series of UVES spectra (Fig.~\ref{fig:transient_shell_BESOS_UVES}, two right panels)---starting at MJD $60238$ ($\phi=0.85$, fourth UVES spectrum chronologically), a sharp absorption component offset by $\sim+55$\,{\kms} from the line center appeared in \spec{He}{i}{5015}. At the next epoch, MJD $60245$ ($\phi=0.93$), the absorption got stronger and shifted to $\sim+40$\,{\kms}. The shell became weaker again at MJD $60256$ ($\phi=0.05$), and shifted to $\sim+50$\,{\kms}. The PUCHEROS and UVES transient events happened during a similar $V>R$ phase, but two entire cycles apart. The remaining instances of less prominent transient shell absorption were identified in CHIRON and BeSS spectra taken during two successive orbits around MJD 61000, while the disk was in $V<R$ phase ($\phi>0.82$ and $\phi<0.08$, Fig.~\ref{fig:transient_shell_overplot}). 

In addition to the transient absorption, the first PUCHEROS epoch ($\phi=0.94$) also shows an emission peak in \spec{He}{i}{6678}, which is offset by $\sim-55$\,{\kms} from the line center. Similarly, \spec{He}{i}{5015} shows an emission peak, offset by $\sim-25$\,{\kms}, at MJD~$60187$ ($\phi=0.27$). While these could be a part of the same variability pattern caused by the companion, transient emission appears to be much rarer than absorption and is not seen in any of the other spectra in our data set.

\begin{figure}[t]
\begin{center}
\includegraphics[width=\hsize]{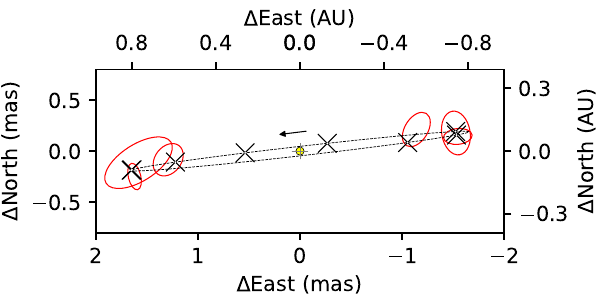}%
\end{center}
\caption[xx]{\label{fig:astrometric_orbit}Astrometric orbit (dashed line) with the primary Be star component at the center ($+$ symbol), and the ascending node on the west side of the orbit. The calculated positions on the orbit for the observation epochs ($\times$ symbols) coincide with the astrometric measurements (error ellipses corresponding to 3$\sigma$ uncertainties). The two epochs with no detections do not have an error ellipse associated with the calculated positions. The arrow shows the direction of motion for the part of the orbit when the sdOB is behind the Be star, barely avoiding being eclipsed.
}
\end{figure}

\section{\label{sec:orbit}Orbital analysis}

We began the orbital analysis with the extraction of astrometric information from the GRAVITY interferometry, followed by a purely astrometric orbital fit. This resulted in the first estimates of orbital parameters such as the period, angular semimajor axis, and inclination. With the knowledge of the period, we were then able to identify spectral lines in our spectroscopic dataset which trace the orbits of both the Be star and the sdOB companion. Combining the astrometric and spectroscopic dataset then resulted in a full, 3-dimensional orbital solution, with model-independent dynamical masses and dynamical parallax.

\subsection{\label{sec:astrometry_fitting}Astrometric orbital solution from interferometric data}

\begin{figure}[th]
\centering

\begin{minipage}{0.24\textwidth}
    \centering
    \includegraphics[width=\linewidth]{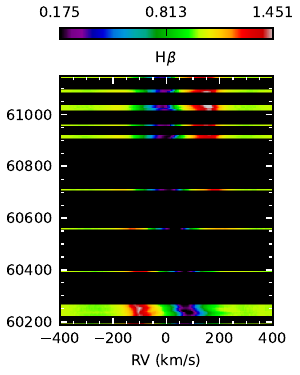}
\end{minipage}
\hfill
\begin{minipage}{0.24\textwidth}
    \centering
    \includegraphics[width=\linewidth]{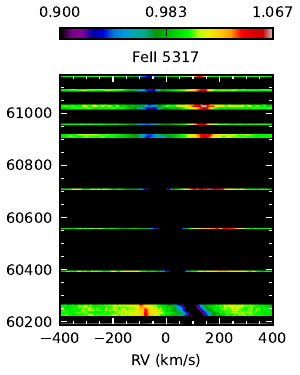}
\end{minipage}

\caption{Time series of spectra plotted in color representation showing the latest transition from $V>R$ to $V<R$ in {\Hb} and \spec{Fe}{ii}{5317}.}
\label{fig:dynspec_vr_longterm}
\end{figure}

For the analysis of individual interferometric snapshots, we employed the Python package PMOIRED \citep{2022SPIE12183E..1NM}. This included the fitting of geometrical models representing the binary system, grid searching for the relative astrometric position and flux ratio of the two binary components (with data rebinned to $R\sim200$), and a bootstrap resampling algorithm to determine uncertainties of the final fitted parameters. 

Both the Be star and the sdOB companion are expected to be unresolved, with the spectrophotometric estimate for the uniform disk (UD) diameter of the Be star in the $K$ band being $0.1105\pm0.0042$\,mas \citep{2016A&A...589A.112C, 2017yCat.2346....0B}. This is consistent with the calibrated |V| remaining close to $1.0$ at all baselines. Attempts to keep the Be star diameter as a free parameter in the binary model resulted in poorly constrained values below $0.5$\,mas, so we fixed it at $0.11$\,mas to avoid overfitting. For the sdOB companion, we considered an unresolved point source. The only free parameters in the binary model were thus the angular separation ($\rho$) and position angle (PA) of the secondary sdOB relative to the Be star, and the flux ratio $f=f_{\rm sdOB}/f_{\rm Be}$. 

For the first and also best-quality epoch -- MJD $59877$ -- the fit resulted in both precise astrometry and the flux ratio ($f_{\rm sdOB}/f_{\rm Be}=0.0365\pm0.010$). For five other epochs, $f$ converged to values consistent with the first epoch, but with larger uncertainties, due to being highly correlated with the angular separation in a certain direction. As we had a good estimate of $f$ from the first epoch, we fixed it at the same value, which helped to pin down the relative astrometry without any correlation. Since we do not expect any significant short term variability in the broadband near-IR fluxes of the components, fixing the flux at a constant value should be well justified. In the two remaining epochs, we could not detect any signal from the companion, indicating that the separation between the components was too small to be resolved by our data. The final set of six astrometric points and two upper limits on the separation revealed an edge-on, circular orbit with about a three-month long period (Table~\ref{tab:astrometry}).

We used the Python code spinOS\footnote{\url{https://github.com/matthiasfabry/spinOS}} \citep{2021A&A...651A.119F,2021ascl.soft02001F} to fit the astrometric orbit. At first, we included the two `upper limit' data points in the fit, but we ended up discarding them entirely, as they had no significant effect on the resulting orbital solution. Still, the two non-detections provided an important consistency check for the orbital solution - the predicted angular separations for those two epochs are $\sim0.5$ and $\sim0.3$\,mas, i.e. both significantly lower than at the other epochs and below what we expect to be able to resolve in our data. The resulting orbital parameters from the purely astrometric fit remained similar in the final, combined orbital solution (Table~\ref{tab:orbital_solution}, Sect.~\ref{sec:combined_solution}). The resulting inclination is high at $89.4\pm2.6$ as expected from the shell spectrum. However, as no eclipses were observed in TESS data, we can derive how much the inclination has to differ from $90^{\circ}$, based on estimated stellar radii, and the distance. This is done is Sect.~\ref{sec:TESS}, resulting in  $i$ of at most $88.4^{\circ}$. Fig.~\ref{fig:astrometric_orbit} shows the astrometric orbit with inclination fixed at $88.4^{\circ}$.

\begin{table}[]
\caption[xx]{\label{tab:orbital_solution} Combined orbital solution.}
\begin{center}
\begin{tabular}{llc}
\hline\hline
Parameter & &  \\
\hline
$P$ & [d] & $87.735\pm0.062$ \\
$T_0$\tablefootmark{a} & [MJD] & $59988.41\pm0.57$ \\
$e$ & & $0$ \\
$a''$ & [mas] & $1.666\pm0.026$ \\
$i$ & [$^\circ$] & $89.4\pm2.6$\tablefootmark{b} \\
$\Omega$ & [$^\circ$] & $276.5\pm1.3$ \\
$\omega_{\rm Be}$\tablefootmark{a} & [$^\circ$] & $90$ \\
$K_{\rm Be}$ & [\kms] & $8.8\pm2.0$ \\
$K_{\rm sdOB}$ & [\kms] & $91.6\pm2.2$ \\
$V_{\rm sys}$ & [\kms] & $15.9\pm1.5$ \\
\hline
$a$ & [AU] & $0.810\pm0.024$ \\
$q=M_{\rm sdOB}/M_{\rm Be}$ & & $0.096\pm0.022$ \\
$M_{\rm total}$ & [\Msun] & $9.20\pm0.83$ \\
$M_{\rm Be}$ & [\Msun] & $8.39\pm0.67$ \\
$M_{\rm sdOB}$ & [\Msun] & $0.81\pm0.22$ \\
$d$ & [pc] & $486\pm16$ \\
\hline
rms astrometry & [$\mu$as] & $59$ \\
rms RV$_{\rm Be}$ & [\kms] & $4.5$ \\
rms RV$_{\rm sdOB}$ & [\kms] & $11.4$ \\
\hline
\end{tabular}
\end{center}
\tablefoot{
$P$ is the orbital period, $T_0$ is the epoch of the inferior conjunction (when the sdOB passes in front of the Be star), $e$ is the eccentricity, $a''$ is the angular semimajor axis of the orbit, $i$ is the orbital inclination, $\Omega$ is the longitude of the ascending node, $\omega$ is the longitude of the periastron, $K$ are the velocity semi-amplitudes, $V_{\rm sys}$ is the systemic velocity, $a$ is the orbital semimajor axis, $q=M_{\rm sdOB} / M_{\rm Be}$ is the mass ratio, $M_\mathrm{total}$ is the total mass, and $M_{\rm Be}$ and $M_{\rm sdOB}$ are the masses of the two components. ``rms'' stands for the root mean square residuals of specified datasets. \\
\tablefoottext{a}{Reference epoch corresponding to inferior conjunction -- when the companion passes in front of the Be star -- for the listed parameters.}
\\
\tablefoottext{b}{Lack of eclipses limit $i$ to $\le88.4^{\circ}$. Adopting $88.4^{\circ}$ rather than $89.4^{\circ}$ has negligible influence on the calculated masses and the distance.}
}
\end{table}

\subsection{Radial velocities from CHIRON spectra}

\begin{figure*}[th]
\centering

\begin{minipage}{0.24\textwidth}
    \centering
    \includegraphics[width=\linewidth]{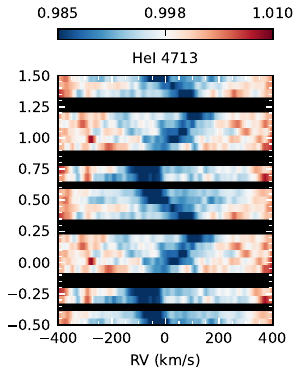}
\end{minipage}
\hfill
\begin{minipage}{0.24\textwidth}
    \centering
    \includegraphics[width=\linewidth]{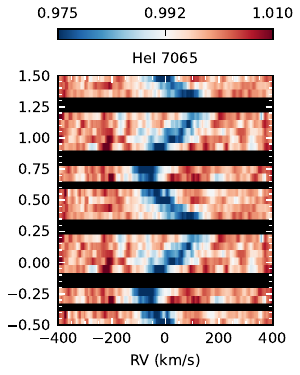}
\end{minipage}
\hfill
\begin{minipage}{0.24\textwidth}
    \centering
    \includegraphics[width=\linewidth]{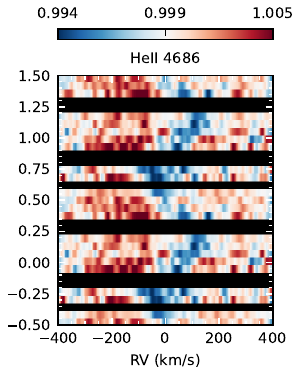}
\end{minipage}
\hfill
\begin{minipage}{0.24\textwidth}
    \centering
    \includegraphics[width=\linewidth]{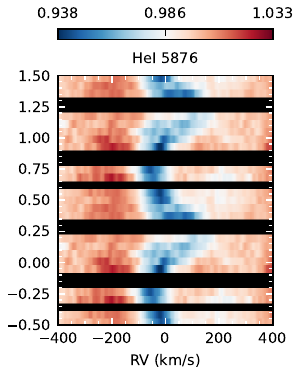}
\end{minipage}

\caption{The full set of 19 CHIRON spectra phased with the orbital period and showing the RV variations of the sdOB companion in \spec{He}{i}{7065} and \spec{He}{ii}{4686}. The y axis shows the orbital phase, and the color bar the normalized flux scale. The shape of the \spec{He}{i}{5876} variation is disturbed by the shell absorption from the disk.}
\label{fig:dynspec_sdO}
\end{figure*}

With the knowledge of the astrometric orbit, we can use the orbital period to plot orbital-phased spectra from intervals covering a slowly evolving part of the $V/R$ cycle, in order to visually detect any periodic RV changes in different spectral lines. The most homogeneous dataset for this purpose are the CHIRON spectra, the last 17 of which were taken around the minimum of the latest $V<R$ phase, while the main shell absorption was redshifted and mostly static (from MJD $60904$ onwards). To illustrate this, we show the latest transition from $V>R$ to $V<R$ as seen in {\Hb} and \spec{Fe}{II}{5018} in Fig.~\ref{fig:dynspec_vr_longterm}. 

\begin{figure*}[th]
\centering

\begin{minipage}{0.24\textwidth}
    \centering
    \includegraphics[width=\linewidth]{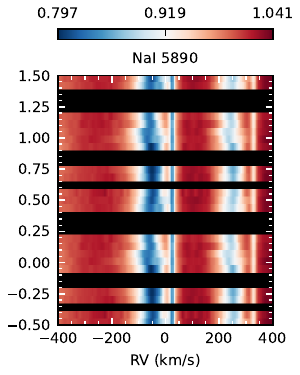}
\end{minipage}
\hfill
\begin{minipage}{0.24\textwidth}
    \centering
    \includegraphics[width=\linewidth]{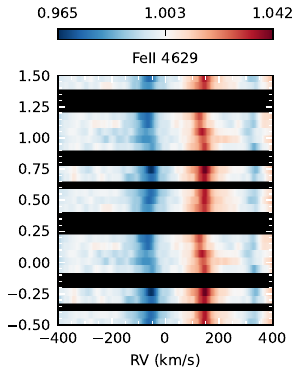}
\end{minipage}
\hfill
\begin{minipage}{0.24\textwidth}
    \centering
    \includegraphics[width=\linewidth]{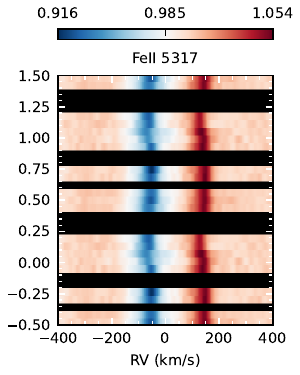}
\end{minipage}
\hfill
\begin{minipage}{0.24\textwidth}
    \centering
    \includegraphics[width=\linewidth]{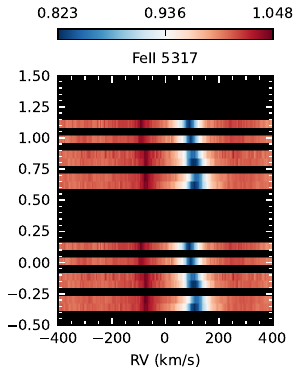}
\end{minipage}

\caption{Same as Fig.~\ref{fig:dynspec_sdO}, but showing spectral lines tracing the orbital motion of the Be star. \spec{Na}{i}{5890}, \spec{Fe}{ii}{4629}, and \spec{Fe}{ii}{5317} (left three panels) were extracted from CHIRON, excluding the first two epochs (see text), while the right panel shows \spec{Fe}{ii}{5317} from our seven UVES spectra. The Be star RVs in the CHIRON dataset obtained during $V<R$ phase of the large scale $V/R$ cycle oscillate around $\sim -60$, $\sim -32$, and $\sim -71$\,\kms, respectively, for the three lines. The UVES \spec{Fe}{ii}{5317} was taken while $V>R$, with the shell absorption shifted to $\sim +100$\,\kms. Static interstellar absorption is seen in \spec{Na}{i}{5890}, and there is an emission component in the \ion{Fe}{ii} lines, shifted to the opposite side than the shell absorption, but also tracing the orbital motion of the Be star.}
\label{fig:dynspec_Be}
\end{figure*}

\begin{figure}[t]
\begin{center}
\includegraphics[width=\hsize]{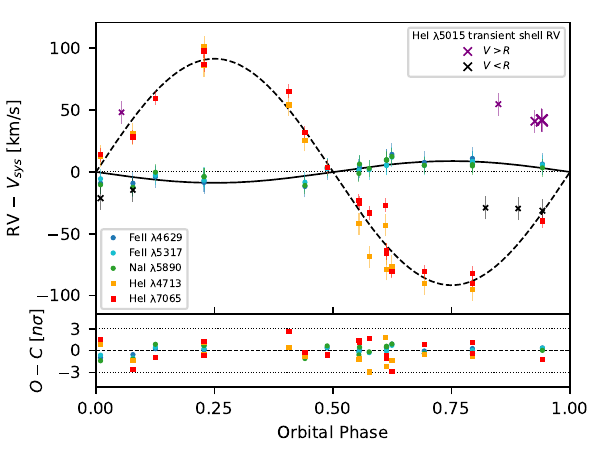}%
\end{center}
\caption[xx]{\label{fig:RV_curve} RV curve of the Be star (solid) and the sdOB companion (dashed), overplotted with RV measurements from lines specified in the legend (lower left), and showing the residuals in the bottom panel. The two sets of transient shell RVs measured in \spec{He}{i}{5015} show little variation, and the difference between the two sets comes from the distinct phases of the large-scale disk oscillation. The symbol sizes for the transient shell RVs are proportional to the depth of the shell absorption. 
}
\end{figure}

The \ion{He}{i} lines where we clearly detected the photospheric signature of the sdOB companion (see Fig.~\ref{fig:dynspec_sdO}) include \spec{He}{i}{4713} (absorption $\sim 1.3\%$ below surrounding continuum), \spec{He}{i}{7065} ($\sim2.5$\%), and \spec{He}{i}{5876} ($\sim5$\% but uncertain due to complex surrounding continuum). A faint signal from \spec{He}{ii}{4686} ($\sim0.5$\%) was clearly detected as well (Fig.~\ref{fig:dynspec_sdO}). To measure the sdOB RVs, we selected \spec{He}{i}{4713} and \spec{He}{i}{7065}, as those two lines have the most symmetric and well defined absorption profiles that maintain similar shape throughout the orbital cycle. Fitting a single Gaussian to the absorption resulted in 34 unique RV measurements (Table~\ref{tab:RVs_sdOB}), with a few lower-confidence measurements from higher-noise spectra discarded. Since we detect the photospheric signature of the sdOB directly, the RVs are not influenced by the $V/R$ cycle, and we can include the earliest two CHIRON spectra taken when $V\approx R$. Unfortunately, both \spec{He}{i}{4713} and \spec{He}{i}{7065} are either missing from the spectral coverage of the remaining spectra (UVES), or the quality is insufficient for reliable RV measurement (low resolution in XSHOOTER, and low signal-to-noise in BeSS spectra).

The RVs of the Be star are difficult to measure directly due the rotationally broadened photospheric lines, with the next best option being to use lines originating in the inner parts of the Be disk as a proxy for the Be star itself. This can be done by tracing the wide emission wings of H$\alpha$\,--\,a method successfully applied to at least a dozen Be stars \citep[e.g.,][]{2020ApJ...898L..44G}\,--\,but the edge-on orientation of AN Col offers a simpler way of using the narrow disk-forming shell-absorption lines. In the CHIRON dataset, we identified signals in three narrow spectral lines that clearly trace the Be star -- \spec{Fe}{ii}{4629}, \spec{Fe}{ii}{5317}, and \spec{Na}{i}{5890} (D2 line, Fig.~\ref{fig:dynspec_Be}) -- and we measured their RVs in the same way as for the companion, resulting in 50 unique RV measurements for the Be star (using the last 17 CHIRON spectra only, Table~\ref{tab:RVs_Be}). Before merging into a single RV dataset, the RVs from each line were corrected for their individual systemic velocity obtained from individual SB2 orbital fits, while the actual systemic velocity is obtained from the photospheric RVs of the sdOB star. 

While the set of the last 17 CHIRON spectra is the most homogeneous and obtained in the most stable phase of the $V/R$ cycle (Fig.~\ref{fig:dynspec_vr_longterm}), there is a possible slow increasing trend in the shell line RVs used for the Be star, becoming apparent especially in the final spectra of the series. This is almost certainly due to the $V/R$ cycle beginning the next phase of increasing $V/R$ ratio and shell core RVs, together with the likely commencement of the next triple-peaked {\Ha} stage. However, the magnitude of this trend is estimated to be below the non-systematic scatter of the measurements due to photon-noise and quality of the spectra, so that adding free parameters to remove such a trend would not lead to more confidence in the solution.

As for the errors of the measured RVs, we recall here that the CHIRON spectra were filtered with a $0.3\,\AA$-wide Gaussian---without this filtering, the strong pixel-to-pixel variations would dominate the fits, resulting in Gaussian FWHMs of 1 pixel and completely arbitrary RV measurements. Since the centering of a Gaussian can easily be done to at least 10 times better than the FWHM, we initially assume a measurement error of 0.03\,{\AA} for the RVs, or about 1.9\,{\kms} for \spec{Fe}{ii}{4629} and \spec{He}{i}{4713}, and 1.7\,{\kms}, 1.5\,{\kms}, and 1.2\,{\kms}, respectively, for \spec{Fe}{ii}{5317}, \spec{Na}{i}{5890}, and \spec{He}{i}{7065}. A double-lined spectroscopic orbital fit to these RVs resulted in reduced $\chi^2$ of 22.0, which means that for the final, combined solution, in which the astrometric and RV residuals are balanced to each result in reduced $\chi^2 \sim 1$ (see below), the initial RV errors were multiplied by a factor of $\sqrt{22} = 4.7$ (Fig.~\ref{fig:RV_curve} shows the measured RVs with scaled errors).

\subsection{Combined orbital solution} \label{sec:combined_solution}

For the 3-dimensional orbital solution combining astrometry and RVs, we again used the spinOS code, and we scaled the bulk of the measurement errors for the two datasets so that the reduced $\chi^2$ resulting from the orbital fit was $\sim 1$ for each. The final orbital solution is given in Table~\ref{tab:orbital_solution}, while the astrometric orbit and the RV curves are plotted in Figs.~\ref{fig:astrometric_orbit} and \ref{fig:RV_curve}, respectively.

The derived orbital distance shows discrepancy with the distances derived from Gaia DR3 for the inner binary as well as the wide B component (Appendix~\ref{appendix:gaia}, Table~\ref{tab:distances}). Considering an orbital solution based only on the Gaia distance, astrometry, and sdOB RVs (ignoring the Be star RVs measured via inner disk shell proxy), we are unable to obtain realistic dynamical masses, and a larger distance is clearly required. While the Gaia parallax for the inner Be + sdOB binary is likely biased by its binary nature, the cause of the discrepancy remains unclear regarding the wide B component, which is expected to be gravitationally bound \citep{2022A&A...657A...7K}. The one hint that the Gaia astrometric solution for the B component could be biased is its slightly elevated Renormalized Unit Weight Error (RUWE) of 1.32.

\begin{figure*}[th]
\centering

\begin{minipage}{0.49\textwidth}
    \centering
    \includegraphics[width=\linewidth]{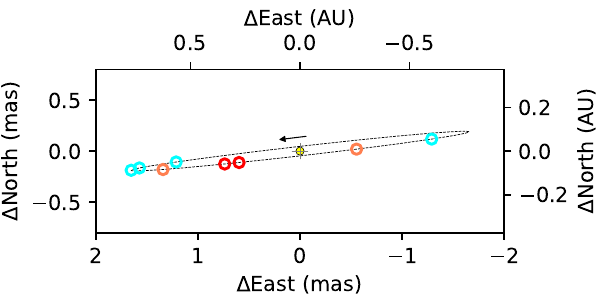}
\end{minipage}
\hfill
\begin{minipage}{0.49\textwidth}
    \centering
    \includegraphics[width=\linewidth]{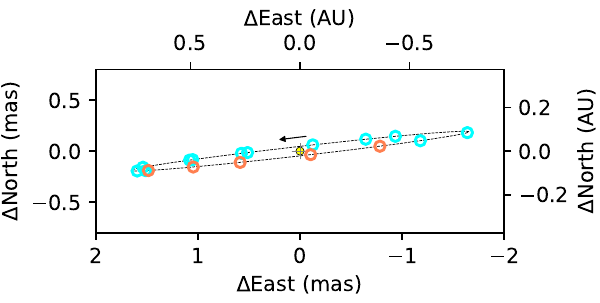}
\end{minipage}

\caption{
The astrometric orbit with epochs of CHIRON and UVES spectra shown as empty circles: epochs are distinguished by the occurence of the transient shell absorption (strong absorption in red, weaker in orange, and none in cyan). The left panel shows UVES and PUCHEROS  spectra obtained during $V>R$ phase, and the right panel shows CHIRON and BeSS spectra during $V<R$ phase; given the direction of the orbital motion and assuming that the disk is coplanar, the denser part of the disk material is East of the Be star when $V>R$, and West of it when $V<R$. The transient shell absorption occurs when the sdOB companion passes in front of the Be star, i.e. around inferior conjunction. The two strong transient shell phases occurred 5--6 days before the actual conjunction. 
}
\label{fig:orbit_transient_shell}
\end{figure*}

\section{\label{sec:discussion}Discussion}

\subsection{Nature and evolutionary origin}

AN~Col is a post-mass-transfer binary system consisting of a rapidly rotating classical $8.4$\,{\Msun} Be ($\sim$ B2 Ve, $T_{\rm eff} \sim 22$\,kK) star and a $0.8$\,{\Msun} stripped subdwarf OB-type companion on a $87.7$-d circular orbit (Table~\ref{tab:orbital_solution}). The orbital inclination is close to $90^{\circ}$, but the absence of eclipses in TESS photometry requires the inclination to be at most $88.4^{\circ}$. The orbital and component properties are similar to the $\sim 15$ other well-characterized Be + sdOB binaries \citep{2015A&A...577A..51M, 2021AJ....161..248W, 2024ApJ...962...70K}, as well as to the several known examples of their direct predecessors---the initial mass donors in Be + bloated pre-sdOB binaries \citep{2020A&A...639L...6S, 2025A&A...694A.208K, 2025A&A...694A.172R}.

The $T_{\rm eff}$ of the sdOB companion can be estimated from the detected photospheric lines. \ion{He}{i} photospheric absorption is expected to peak at 20-22\, kK, but remains remains strong even up to $30$\,kK before getting weaker above \citep{1998A&A...330..306L}. Ratios to the detected \spec{He}{ii}{4686} line (see Fig.~\ref{fig:dynspec_sdO}) are then particularly sensitive to $T_{\rm eff}$ in the relevant range. The depression of the \spec{He}{ii}{4686} line is about 0.5\% below the continuum, while the nearby \spec{He}{i}{4713} is about 1.3\% below the continuum --- Judging from measured line strength ratios in synthetic spectra at our disposal (\citealt{1992IAUS..149..225K, 2011ascl.soft09022H}, see Sect.~3 of \citealt{2013MNRAS.429..177R} for more details), this suggests an effective temperature around 30\,kK, and a classification of about B0 or O9.5 for the subdwarf (since both lines are just a few nm apart, any systematic deviations from a purely photospheric early B-type star SED, possibly introduced by disk emission, will be very small and not affect this estimate). In Sect.~\ref{appendix:gaia}, we also made a rough estimate that the sdOB radius should be around $0.75$\,{\Rsun}. As for the rotational broadening of these photospheric lines, the \spec{He}{i}{4713} line FWHM is 80\,\kms, indicating a rather slow rotational velocity --- This is in line with what is typically observed for sdOB companions and their bloated predecessors \citep{2023AJ....165..203W, 2025A&A...694A.172R}.

We investigated the possible initial configuration and mass transfer history using the method of \citet[][their Eqs. B.3 and B.4]{2025ApJ...990L..51L}. Assuming case B mass transfer, the strict limit on the minimum mass transfer efficiency $\beta_{\rm min}$ is $54$\%, while the best-guess minimum efficiency is $79$\%. These values are at the higher end, but within the range of what was obtained for 16 other Be star binaries by \citet{2025ApJ...990L..51L}, adding further evidence that $\beta_{\rm min}$ is higher than generally expected. Under the same assumptions adopted by \citet{2025ApJ...990L..51L}, i.e. $10$\% angular momentum loss to a circumbinary disk whose radius is twice that of the orbital semimajor axis, and taking the best-guess $\beta_{\rm min}$ of $79$\%, the initial masses for the Be-star and sdOB progenitors are similar at $5.0\,\Msun$, the initial orbital period is $2.6$ days, and the initial semimajor axis $0.10$\,AU.

\subsection{Density oscillation in the disk and the $V/R$ cycle}

The observed long-term cyclic spectral variations come from a one-armed density pattern slowly precessing around the star \citep{1991PASJ...43...75O}. Since AN~Col is a shell-Be star (i.e., seen edge-on), the same disk cycle manifests itself not only in the emission spectrum, but also in the shell absorption spectrum, which presents cyclic shifts in its prominence and RV offset (Fig.~\ref{fig:EW_RV_VR_stacked}). The Be disk and the binary orbit in AN~Col are co-aligned and share the same sense of rotation, with the approaching side of the disk in the East, as revealed by GRAVITY spectrointerferometry. The signature of the $V/R$ cycle is stronger when the dense part approaches the observer on the East side than when it is receding on the other side, which could be due to shielding of the leading dense material by more tenuous parts.

The disk oscillation cycle in AN~Col is strikingly similar to that observed in the early-type shell-Be star $\zeta$~Tau. The three full $V/R$ cycles that AN~Col went through from 2010 to 2026 appear to match almost exactly the three cycles seen in $\zeta$~Tau between 1997 and 2009 \citep{2009A&A...504..929S}, including the $V/R$ ratio oscillating around a value which was not unity, as well as the onset of a triple-peaked {\Ha} phase following each minimum in $V/R$. Also like in $\zeta$~Tau, AN~Col shows a phase lag in the $V/R$ (and shell absorption RV) between different lines, which come from the spiral shape of the oscillation together with the distinct formation regions within the disk of the different lines. The fact that in AN~Col the $V/R$ phase of Br$\gamma$ appears to trail H$\alpha$ rather than precede it could be caused by different helicity of the spiral arm than in $\zeta$~Tau \citep[cf. Fig.~5 of][]{2009A&A...504..915C}. In the near-IR, our AN~Col spectra are of somewhat higher resolution and reveal sharper shell absorption at some epochs than observed in $\zeta$~Tau, but otherwise the spectrum is the same, even including emission components occasionally observed in the \ion{C}{i} multiplet 1 \citep[][their Fig.~8]{2009A&A...504..929S}, which appear to be disconnected from both the $V/R$ cycle, and the binary orbit.

\citet{2009A&A...504..915C} and \citet{2015A&A...576A.112E} successfully reproduced most of the spectroscopic and spectro-interferometric variations in $\zeta$~Tau using the VDD model \citep{2006ApJ...639.1081C, 2008ApJ...684.1374C} together with the precessing one-armed density oscillation model \citep{1997A&A...318..548O, 1992A&A...265L..45P}. Given the extent of similarities of the spectral variations seen in AN~Col, we can confidently conclude that the underlying mechanism is the same in AN~Col as it is in $\zeta$~Tau. Potential detailed modeling of the density oscillation is deferred for a later study, but better temporal coverage of the spectroscopic data will be needed to gain significant new insights into the origin of this phenomenon.

Like AN~Col, $\zeta$~Tau is also a post-mass-transfer binary star with a close companion, although it is a white dwarf rather than sdOB star \citep{2024A&A...688A.181N, 2026A&A...707A.334N}. The strong, long-term $V/R$ cycle, however, does not generally appear to be connected to binarity. Much weaker, orbital phase-locked, double-armed density patterns may be excited by the tidal influence of a companion \citep{2007ASPC..361..274S, 2023Galax..11...83M}, but if present, this variation is overwhelmed by the stronger, long-term $V/R$ cycle in both AN~Col and $\zeta$~Tau.

The occurrence of the triple-peaked phase in {\Ha} seems to be exclusive to shell-Be stars \citep[it can be described as forming from `a doubling of the self-absorption in the disk' according to][]{2009A&A...504..929S}, and such profiles seem to only appear when the dense material forming the $V/R$ oscillation is behind the Be star \citep{2006A&A...459..137R}. Triple-peaked and similarly disturbed {\Ha} profiles (see Fig.~\ref{fig:triple_peak}) are also more commonly observed in Be stars with known companions, although no robust statistical study of their occurrence has yet been carried out \citep{2006A&A...459..137R}. As this phase occurs with the dense material behind the Be star, and because the disk particles in an oscillating disk follow ordered, co-precessing elliptical orbits, the triple-peaked phase occurs when the common apastron of the particle orbits is oriented toward the observer. This also happens to be the part of the particle orbits where they are most susceptible to tidal perturbations exerted by the companion. Supporting the tentative conclusion that triple peaked phases occur in binary Be-shell stars only, another very similar Be shell star $48$\,Lib shows a $V/R$ cycle akin to that of AN~Col and $\zeta$\,Tau, but lacks both the triple-peaked phases and any report of a binary companion \citep{2016ApJ...826...81S}.

\subsection{Be disk--sdOB companion interaction}

Transient variations caused by a close companion similar to those in AN~Col have been observed for a number of other Be stars, among them $\varphi$~Per \citep{2000A&A...358..208S, 2001A&A...368..471H}, FY~CMa \citep{2004A&A...427..307R}, HR~2142 \citep{2016ApJ...828...47P}, HD~55606 \citep{2018ApJ...865...76C}, and V658~Car \citep{2025ApJ...995...87D}. In AN~Col, the variations and their origin from outer Be disk are qualitatively most similar to those observed in $\varphi$~Per and modeled by \citet{2001A&A...368..471H}.

The periodic transient shell absorption linked to the binary orbit occurs most prominently in \ion{He}{i} lines, especially \spec{He}{i}{5015} (Figs.~\ref{fig:transient_shell_BESOS_UVES}, \ref{fig:orbit_transient_shell}, and \ref{fig:transient_shell_overplot}). The RVs of the transient shell absorption do not vary significantly as the companion orbits (Sect.~\ref{sec:companion_induced_variability}, Fig.~\ref{fig:RV_curve}), while maintaining a long-term offset consistent with the phase of the $V/R$ cycle. This fact favors the less extended outer parts of the Be disk projected on the Be photosphere as the origin of the transient shell \citep[cf. $\varphi$~Per;][]{2000A&A...358..208S, 2001A&A...368..471H}. 

The situation is different in HR~2142 and HD~55606 where the RV variation of the same shell feature changes by several tens of {\kms} during a single orbit, so that it is thought to originate in extended spiral arms of the Be disk \citep[HD~55606;][]{2018ApJ...865...76C} or in tidal streams connecting the outer disk to the companion and beyond \citep[HR~2142;][]{2016ApJ...828...47P}. In the eclipsing Be + sdOB binary V658~Car \citep{2025ApJ...995...87D}, the origin of the transient shell is still different as it comes from the sdOB when it passes behind the Be star and shines through its disk, resulting in the largest shell RV variations during a single orbit. 

Other than the periodic shell events, the absorption lines of the sdOB companion seem to appear slightly deeper when the sdOB is approaching than when it is receding (Fig.~\ref{fig:dynspec_sdO}). This changing line strength resembles the phenomenon referred to as the Struve-Sahade effect \citep[][and references therein]{1997ApJ...479..408G}. In AN~Col, this could happen if the companion creates something like a wake as it passes through the tenuous outer parts of the Be disk, and in the context of the state-of-the-art hydrodynamical simulations of \citet{2025A&A...698A.309R}, this may be naturally explained by the `bridge' part of the disk extending towards a binary companion.

\section{\label{sec:concl}Conclusions} 

AN~Col is only the second Be + sdOB binary (and only the fourth Be + stripped star binary) with a combined astrometric and SB2 orbital solution resulting in model-independent dynamical masses and geometrical-parallax-independent distance. The model-independent dynamical masses make AN Col an important benchmark for testing binary evolution calculations. Within the sample of well-characterized Be + stripped star binaries, AN~Col confirms the trend that these systems require high mass transfer efficiencies during the previous mass-transferring phase that led to the spin-up of the Be star. 

AN~Col is also another Be star where the presence of a hot stripped companion was revealed by its influence on the outer parts of the Be-star disk. It is also another in a small number of Be-shell stars where additional transient shell absorption occurs as a line-of-sight effect caused to the companion. The properties of the transient shell absorption indicate that they originate in the outer irradiated part of the Be disk, similar to $\varphi$~Per, rather than in an extended spiral arm or a `bridge' connecting the outer disk with the companion like in HR~2142 and HD~55606.

The circumstellar disk of the primary Be star is subject to long-term cyclic oscillations in the form of a propagating one-armed density wave, giving rise to strong $V/R$ ratio and RV variations in emission lines and in shell absorption lines, respectively. These variations happen with a cycle length of $5.1$\,yr with three full $V<R \rightarrow$ Triple peak $\rightarrow V>R \rightarrow V\approx R$ cycles completed over the last $15.6$ years. The variability due to the global disk oscillation bears a remarkably close resemblance to that of $\zeta$~Tau, for which a dedicated oscillation model exists.

Future detailed modeling of the $V/R$ cycle and the Be-companion hinges on continued monitoring with high-resolution spectroscopy and spectro-interferometry. Atmospheric analysis of the $\sim30$\,kK sdOB companion would be particularly compelling given the SB2 status of the binary.

\begin{acknowledgements}
A. C. C. acknowledges support from CNPq (grant 314545/2023-9) and FAPESP (grants 2011/51680-6, 2019/26492-3, 2023/05083-3, 2025/28195-7).
A. S. C. acknowledges financial support from the GSU Second Century Initiative (2CI) Fellowship.
Based on observations collected at the European Southern Observatory under ESO programmes 093.D-0415(A), 60.A-9022(C), 110.23Z7.001, 112.25KN.001, and 112.25KN.002.
This work used the BeSOS Catalogue, operated by the IFA-UV, Chile: \url{http://besos.ifa.uv.cl} and is maintained by Fondecyt No. 11190945.
This work has made use of the BeSS database, operated at LESIA, Observatoire de Meudon, France: \url{http://basebe.obspm.fr}.
This work uses data from the CTIO/SMARTS-GSU 1.5m telescope, which is operated as part of the SMARTS Consortium by RECONS (\url{www.recons.org}) members Todd Henry, Tim Johns, and Sebastian Carrazco Gaxiola. At the telescope, observations were carried out by Roberto Aviles, Ivonne Toro Bertolla, Emma Galligan, Todd Henry, Tim Johns, and Aman Kar.
This research has made use of the Jean-Marie Mariotti Center \texttt{Aspro} service\footnote{Available at http://www.jmmc.fr/aspro} \citep{2016SPIE.9907E..11B}. This research has made use of the Jean-Marie Mariotti Center \texttt{OIFits Explorer} service\footnote{Available at http://www.jmmc.fr/oifitsexplorer}. This research has made use of the Jean-Marie Mariotti Center \texttt{SearchCal} service, which involves the JSDC and JMDC catalogues\footnote{Available at https://www.jmmc.fr/searchcal} \citep{2016A&A...589A.112C}. 
This work has made use of data from the European Space Agency (ESA) mission
{\it Gaia} (\url{https://www.cosmos.esa.int/gaia}), processed by the {\it Gaia}
Data Processing and Analysis Consortium (DPAC,
\url{https://www.cosmos.esa.int/web/gaia/dpac/consortium}). Funding for the DPAC
has been provided by national institutions, in particular the institutions
participating in the {\it Gaia} Multilateral Agreement.
This paper includes data collected with the TESS mission, obtained from the MAST data archive at the Space Telescope Science Institute (STScI). Funding for the TESS mission is provided by the NASA Explorer Program. STScI is operated by the Association of Universities for Research in Astronomy, Inc., under NASA contract NAS 5–26555.
This research has made use of the SIMBAD database, CDS, Strasbourg Astronomical Observatory, France \citep{2000A&AS..143....9W}. This research has made use of the VizieR catalogue access tool, CDS, Strasbourg Astronomical Observatory, France (DOI : 10.26093/cds/vizier) \citep{2000A&AS..143...23O}.

\end{acknowledgements}

\bibliographystyle{aa} 
\bibliography{ANCol} 

\begin{appendix}

\section{Tabulated measurements of astrometric positions and RVs}\label{appendix:tables}

\renewcommand{\arraystretch}{1.2}
\begin{table*}[h!]
\caption{Relative astrometric positions of the AN Col binary system determined from GRAVITY data.}         
\label{tab:astrometry}      
\centering          
\begin{tabular}{l c c c c c c c c c c} 
\hline\hline       
Date & MJD & $\rho$ & PA & $\Delta$RA & $\Delta$DEC & $\sigma$-$a$ & $\sigma$-$b$ & $\sigma$-PA & Baselines & $f=f_{\rm sdOB}/f_{\rm Be}$ \\ 
 &  & & [mas]  & [$^\circ$] & [mas] & [mas] & [mas] & [mas] & [$^\circ$] &  \\
\hline
2022~Oct~24 & 59877.280 & 1.641 & 98.739 & 1.622 & -0.249 & 0.0442 & 0.0204 & 13.2 & A0-G2-J2-J3 & 0.0365$\pm$0.0010\\
2022~Dec~14 & 59928.123 & 1.539 & 276.627 & -1.528 & 0.178 & 0.0721 & 0.0464 & 9.7 & A0-G1-J2-K0 & 0.0365\\
2023~Oct~01 & 60219.298 & 1.295 & 93.746 & 1.292 & -0.085 & 0.0581 & 0.0429 & 142.3 & A0-G1-J2-K0 & 0.0365\\
2023~Nov~12 & 60261.225 & 1.160 & 280.528 & -1.141 & 0.212 & 0.0618 & 0.0377 & 148.3 & A0-G2-J2-J3 & 0.0365\\
2023~Nov~19 & 60268.130 & 1.554 & 275.312 & -1.547 & 0.144 & 0.0469 & 0.0258 & 95.7 & A0-B5-J2-J6 & 0.0365\\
2023~Dec~14 & 60293.178 & -- & -- & -- & -- & -- & -- & -- & A0-G2-J2-J3 & --\\
2023~Dec~21 & 60300.062 & -- & -- & -- & -- & -- & -- & -- & A0-B5-J2-J6 & --\\
2024~Jan~07 & 60317.061 & 1.587 & 94.108 & 1.583 & -0.114 & 0.1255 & 0.0597 & 122.6 & A0-B5-J2-J6 & 0.0365\\
\hline
\end{tabular}
\tablefoot{$\rho$ is the angular separation between the components, PA is the position angle (from north to east), $\Delta$RA and $\Delta$DEC are the Ab coordinates relative to Aa in the east and north direction, respectively, $\sigma$-$a$ and $\sigma$-$b$ are the major and minor axes of the error ellipse, respectively, $\sigma$-PA is the position angle of the error ellipse (from north to east), and $f=f_{\rm sdOB}/f_{\rm Be}$ is the flux ratio.\\
}
\end{table*}

\begin{table}[ht]
\centering
\caption{RVs for the sdOB from photospheric absorption}
\label{tab:RVs_sdOB}
\begin{tabular}{lcc}
\hline
MJD & \spec{He}{i}{4713} & \spec{He}{i}{7065} \\
&
[\kms] & [\kms] \\
\hline
60638.264 & 70.1 & 81.1 \\
60656.205 & -27.2 & -10.9 \\
60904.437 & 41.5 & 47.6 \\
60914.429 & -26.0 & -7.1 \\
60916.330 & -52.7 & -17.3 \\
60954.336 & 28.8 & 29.8 \\
60960.336 & 46.6 & 44.2 \\
61014.265 & -74.3 & -64.8 \\
61023.171 & -79.4 & -74.2 \\
61023.178 &       & -66.1 \\
61036.128 &       & -23.7 \\
61084.020 &       & 19.6 \\
61090.012 & -26.0 & -9.2 \\
61095.006 & -62.8 & -47.5 \\
61095.013 & -50.1 & -49.6 \\
61096.007 & -60.3 & -64.4 \\
61139.968 &       & 75.6 \\
61148.960 & 101.9 & 113.8 \\
61148.968 & 117.2 & 102.8 \\
\hline
\end{tabular}
\tablefoot{These photospheric RV measurements were not corrected for the systemic velocity of $15.9$\,\kms. The scaled uncertainties used in the combined orbital solution were $8.9$\,{\kms} for \spec{He}{i}{4713}, and $5.6$\,{\kms} for \spec{He}{i}{7065}.}
\end{table}

\begin{table}[ht]
\centering
\caption{RVs for the Be star from narrow shell lines}
\label{tab:RVs_Be}
\begin{tabular}{lccc}
\hline
MJD &
\spec{Fe}{ii}{4629} & \spec{Fe}{ii}{5317} & \spec{Na}{i}{5890} \\
& [\kms] & [\kms] & [\kms] \\
\hline
60904.437 & -11.7 & -8.3 & -10.9 \\
60914.429 & 1.9 & 1.3 & -1.2 \\
60916.330 &      & 1.8 & 2.8 \\
60954.336 & -9.1 & -5.5 & -10.4 \\
60960.336 & -9.1 & -12.3 & -12.4 \\
61014.265 & 7.7 & 5.2 & 5.4 \\
61023.171 & 10.9 & 7.5 & 4.8 \\
61023.178 & 8.4 & 8.6 & 5.9 \\
61036.128 & 6.4 & 5.8 & 3.3 \\
61084.020 & 2.5 & 2.9 & 3.8 \\
61090.012 & 3.8 & 5.8 & 6.4 \\
61095.006 & 9.6 & 5.2 & 9.9 \\
61095.013 & 9.6 & 5.8 & 9.4 \\
61096.007 & 14.2 & 12.0 & 12.5 \\
61139.968 & -4.0 & -2.7 & -0.2 \\
61148.960 & -7.2 & -3.8 & -3.8 \\
61148.968 & -8.5 & -6.6 & -3.8 \\
\hline
\end{tabular}
\tablefoot{RV measurements from each line were corrected for their individually determined systemic velocities, so that they all correspond to $V_{\rm sys} = 0$\,\kms.
The scaled uncertainties used in the combined orbital solution were $8.9$\,{\kms} for \spec{Fe}{ii}{4629}, $8.0$\,{\kms} for \spec{Fe}{ii}{5317}, and $7.0$\,{\kms} for \spec{Na}{i}{5890}.}
\end{table}

The measurements from our eight interferometric observations from GRAVITY are detailed in Table~\ref{tab:astrometry}. The RV measurements for the sdOB and Be components are given in Tables~\ref{tab:RVs_sdOB} and \ref{tab:RVs_Be}, respectively. Empty table cells indicate that the profile of the respective line was considered too noisy to obtain a reliable measurement.

\section{Additional observations}\label{appendix:additional_observations}

\begin{table*}
\caption{Broad-band linear polarimetry of AN\,Col obtained with IAGPOL.}
\centering
\small
\setlength{\tabcolsep}{4pt}
\begin{tabular}{cccccccccc}
\hline\hline
 &
\multicolumn{2}{c}{$B$} &
\multicolumn{2}{c}{$V$} &
\multicolumn{2}{c}{$R$} &
\multicolumn{2}{c}{$I$} \\
\cline{2-3}\cline{4-5}\cline{6-7}\cline{8-9}
MJD &
$P$ &
$\theta$ &
$P$ &
$\theta$ &
$P$ &
$\theta$ &
$P$ &
$\theta$ \\
&
(\%) &
($^\circ$) &
(\%) &
($^\circ$) &
(\%) &
($^\circ$) &
(\%) &
($^\circ$) \\
\hline
60564.238 &
$0.821 \pm 0.034$ &
$0.9 \pm 1.2$ &
$0.761 \pm 0.015$ &
$3.5 \pm 0.6$ &
$0.673 \pm 0.004$ &
$9.49 \pm 0.18$ &
$0.660 \pm 0.006$ &
$6.71 \pm 0.26$ \\

60599.249 &
$0.938 \pm 0.029$ &
$4.7 \pm 0.9$ &
$0.793 \pm 0.012$ &
$6.9 \pm 0.4$ &
$0.669 \pm 0.012$ &
$5.7 \pm 0.5$ &
$0.649 \pm 0.025$ &
$5.3 \pm 1.1$ \\
\hline
\end{tabular}
\tablefoot{
Broad-band linear polarization measurements of AN\,Col in the $BVRI$ bands, with $P$ being the polatization percentage, and $\theta$ being the position angle. The quoted uncertainties are the formal $1\sigma$ errors from the IAGPOL reduction pipeline. The MJD corresponds to the mean epoch of the $BVRI$ observing sequence.
}
\label{tab:pol}
\end{table*}

\subsection{Gaia Astrometry}\label{appendix:gaia}

{
\renewcommand{\arraystretch}{1.2}
\begin{table*}[ht!]
\caption{Measurements from Gaia DR3 and Hipparcos}
\label{tab:distances}
\centering
\begin{tabular}{c c c c c c c}
\hline\hline
Simbad ID & $G$ & PM$_{\rm RA}$ & PM$_{\rm dec}$ & RUWE & Gaia $d$\tablefootmark{a} & Hipparcos $d$\\
 & [mag] & [mas\,yr$^{-1}$] & [mas\,yr$^{-1}$] & & [pc] & [pc] \\
\hline
HD 35165 & $6.11$ & $-0.11\pm0.21$ & $7.58\pm0.22$ & 1.00 & $378.4^{+33.3}_{-28.2}$ & $443\pm113$ \\
HD 35165B & $10.35$ & $0.39\pm0.09$ & $7.99\pm0.09$ & 1.32 & $390.4^{+6.9}_{-6.7}$ & - \\
\hline
\end{tabular}
\tablefoot{
\tablefoottext{a} {Taken from the Alma Luminous Star (ALS) catalogue \citep[][\url{https://als.cab.inta-csic.es/}]{2025MNRAS.543...63P} for the main component, and from \citet{2021AJ....161..147B} for the fainter component B.}
}
\end{table*}
}

Gaia DR3 \citep{2016A&A...595A...1G, 2023A&A...674A...1G} lists as separate entries both the main bright component of the wide AN~Col binary (Gaia DR3 4823162172201408128; the Be + sdOB binary), and the fainter component $\sim2$ arcseconds away (Gaia DR3 4823162172201407360, the B component). The angular separation between the two components is $\sim2.19"$, and the position angle (north through east) is $8.50^{\circ}$, with the reference epoch 2016. Thus, no significant change in the relative position compared to the historic astrometry spanning the years 1922 to 1969 \citep{2019yCat....102026M} is detected. 

AN~Col is not listed as a binary candidate based on proper motion anomaly, but it has two candidate companions based on common proper motion \citep{2022A&A...657A...7K}. Of those two companions, the first one is the B component mentioned above, and the other is a faint $G=18.4$ star separated by $\sim62"$, which is not relevant for this work.

In Table~\ref{tab:distances}, we list the relevant Gaia DR3 data for the two main components \citep[the inner Be + sdOB binary and the wide B companion;][]{2025MNRAS.543...63P, 2021AJ....161..147B}, together with the distance derived from revised Hipparcos parallax \citep{2007A&A...474..653V}. The derived distances of the two components agree within $1\sigma$, while their corrected proper motions agree within $2.4\sigma$. The distance derived from the orbital parallax of the inner Be + sdOB binary ($486\pm16$\,pc, Sect.~\ref{sec:orbit}) shows a discrepancy of $2.3\sigma$ compared to the Gaia distance of the inner component, and $3.1\sigma$ to the Gaia distance of the B component, while agreeing well with the Hipparcos distance.

\subsection{Astrometry of the wide pair from ESO acquisition camera images}\label{appendix:astrometry_from_acquisition_cameras}

Given the angular separation of $2.2"$, and the magnitude difference of 4 between AN~Col and its wide companion, both are typically visible on acquisition cameras of various instruments, even when only the brighter component is actually observed. These acquisition images can be used to recover astrometry of the two wide components, revealing no significant orbital movement compared to the historic astrometric measurements.

We recovered an acquisition camera image for AN~Col taken on ${\rm MJD}53727$ (2005 Dec) with the NACO instrument, which was installed at one of the four 8-m telescopes of the VLT. NACO included an adaptive optics (AO) system, and the images are therefore diffraction-limited and should provide accurate astrometry. We measured a separation of $2.2130\pm0.0014"$ at a position angle of $13.03\pm0.04^{\circ}$ by fitting Gaussians. As for the magnitude difference, we recovered only a lower limit of $\sim3.5$\ in the Ks band due to apparent non-linear response of the detector for the brighter primary component.

The acquisition camera images from the XSHOOTER observations are compatible with the above findings. However, the uncertainties in the recovered astrometric positions are
large due the images being seeing-limited (no AO correction), and we are not able to provide measurements that would enable a reliable detection of a possible orbital motion. As for the acquisition camera images associated with our UVES observations, the wide companion is again clearly visible, but the relative astrometry cannot be measured reliably due to the shadow of the entrance slit in the acquisition images.

\subsection{TESS photometry}\label{sec:TESS}

\begin{figure}[t]
\begin{center}
\includegraphics[width=\hsize]{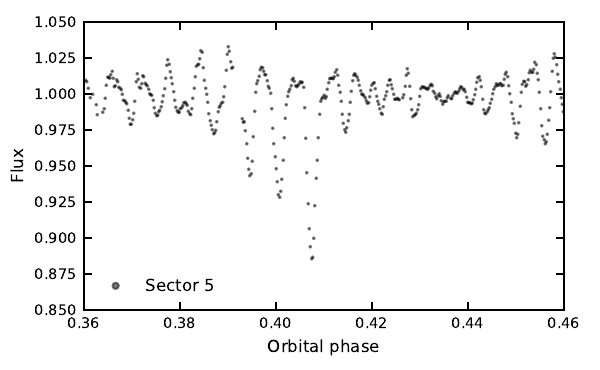}\\[0em]
\includegraphics[width=\hsize]{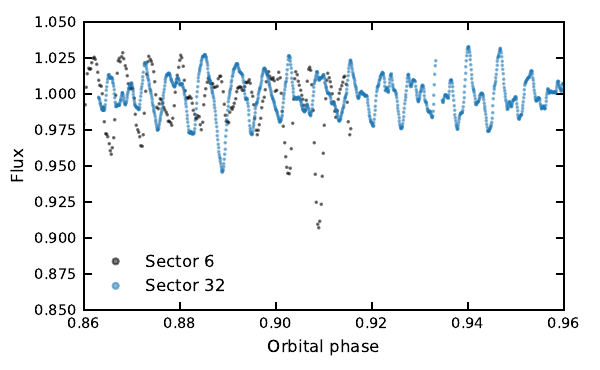}
\end{center}
\caption{TESS light curves from sectors 5 and 6 phased with the orbital ephemeris (Table~\ref{tab:orbital_solution}) and zoomed into the two orbital phases when the observed photometric minima occur (upper and lower plot). The photometric minima around phase 0.91 occurred only during Sector 6 and not during Sector 32. The orbital phase difference between the two photometric minima observed in Sectors 5 and 6 is exactly 0.5.}
\label{fig:TESS}
\end{figure}

AN\,Col was observed during Transiting Exoplanet Survey Satellite \citep[TESS,][]{2015JATIS...1a4003R} sectors 5, 6, 32, and 98, but did not fall on a usable part of the detector during observations of sector 98. Extracted light curve data are available for the first three sectors in the MAST archive\footnote{\url{https://archive.stsci.edu/missions-and-data/tess}}. The appearance of the light curve and frequency spectrum is typical for an early-type Be star \citep{2022AJ....163..226L}, with frequency groups around values of 1.6 and 3.1\,c/d, and a strong, outstanding frequency at 1.84\,c/d and its harmonic at 3.68\,c/d, making this a candidate for the rotation frequency, with peak-to-peak amplitudes of a few tens of mmag and below.

No obvious eclipses are present. However, in sectors 5 and 6 particularly, deep photometric minima of about 12\% and 10\% occur at orbital phases 0.41 and 0.91 (Fig.~\ref{fig:TESS}). These are not shaped like a typical eclipse, but sinusoidal, and appear to show one or two shallower precursor minima. The phase range around 0.91 is also present in sector 32, but this time without any similar signature. For this reason, true geometric eclipses can be ruled out, but the precise time difference of the minima in sectors 5 and 6 of half an orbital period suggests some possible relation to the orbit. Given the complex structure that Be disks attain in close binary configurations, the photometric minima observed in sectors 5 and 6 might be caused by extended outer parts of the Be disk \citep{2025A&A...698A.309R}.

The lack of true geometric eclipses puts limits on the orbital inclination. With a mass of $\sim8.4$\,\Msun\xspace (Sect.~\ref{sec:orbit}), the Be star would be close to spectral type B2 and have a radius of about $\sim4.2$\,\Rsun\xspace according to \citet{1988BAICz..39..329H} \citep[$\sim4.1$\,\Rsun\xspace according to ][]{2013ApJS..208....9P}. The companion, at a flux ratio of 3.7\% in the IR and being slightly hotter, should have about 15 to 20\% of that radius, i.e. around $0.75$\,{\Rsun}. No observable eclipses imply an impact parameter above about 5\,\Rsun. At the given orbital dimensions this means $i<88.4^\circ$, which is well within the error estimated in Table~\ref{tab:orbital_solution} for the inclination.

\section{Additional material on the $V/R$ cycle}\label{appendix:vr_cycle}

\begin{figure}[th]
\centering
\begin{minipage}{0.24\textwidth}
    \centering
    \includegraphics[width=\linewidth]{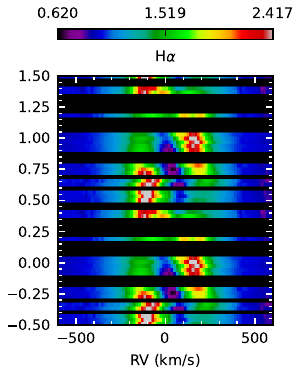}
\end{minipage}
\hfill
\begin{minipage}{0.24\textwidth}
    \centering
    \includegraphics[width=\linewidth]{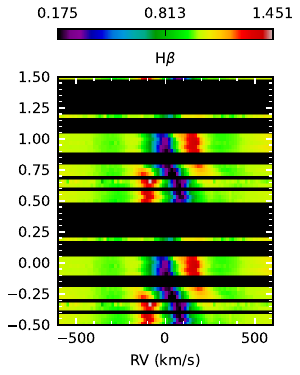}
\end{minipage}
\caption{H$\alpha$ and H$\beta$ spectra phased with the $V/R$ cycle length of 1855\,d.
}
\label{fig:dynspec_vr_phased}
\end{figure}

\begin{figure}[t]
\begin{center}
\includegraphics[width=\hsize]{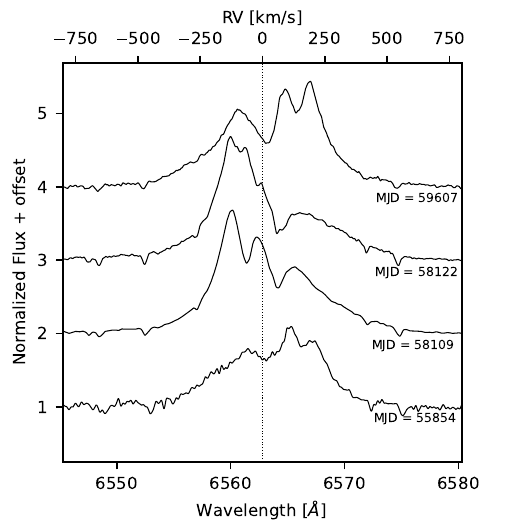}%
\end{center}
\caption{The four spectra in our dataset taken when the V/R ratio was
increasing, corresponding to the triple-peaked phase in the V/R cycle.
In the transitory spectrum from MJD 58122, the middle peak is in the
process of disappearing.}
\label{fig:triple_peak}
\end{figure}

{\Ha} and {\Hb} lines phased with the length of the $V/R$ cycle are shown in Fig.~\ref{fig:dynspec_vr_phased}), while the triple peaked {\Ha} profiles are plotted in Fig.~\ref{fig:triple_peak}. 

\section{Additional material on the companion-induced spectroscopic variations}\label{appendix:additional_transient_shell}

\begin{figure*}[th]
\centering

\begin{minipage}{0.49\textwidth}
    \centering
    \includegraphics[width=\linewidth]{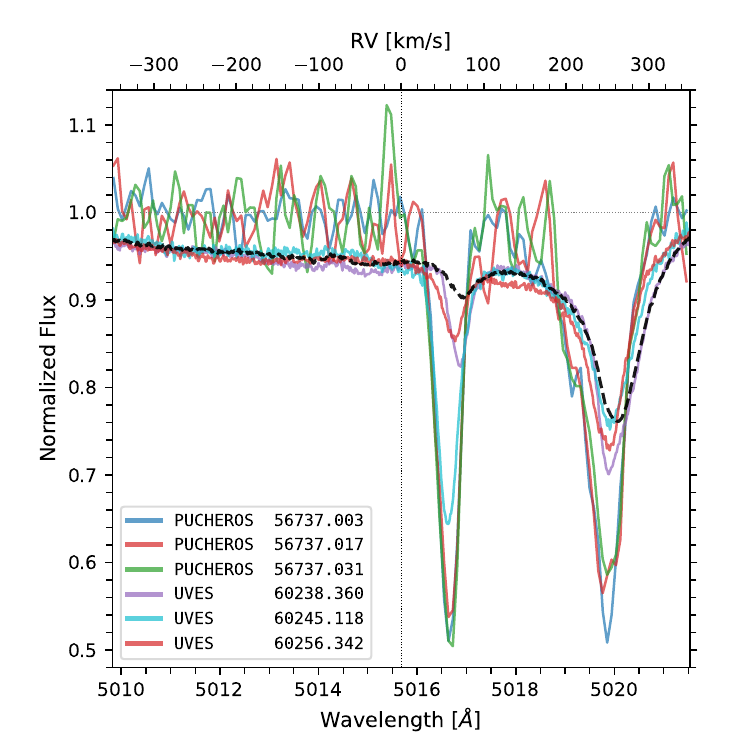}
\end{minipage}
\hfill
\begin{minipage}{0.49\textwidth}
    \centering
    \includegraphics[width=\linewidth]{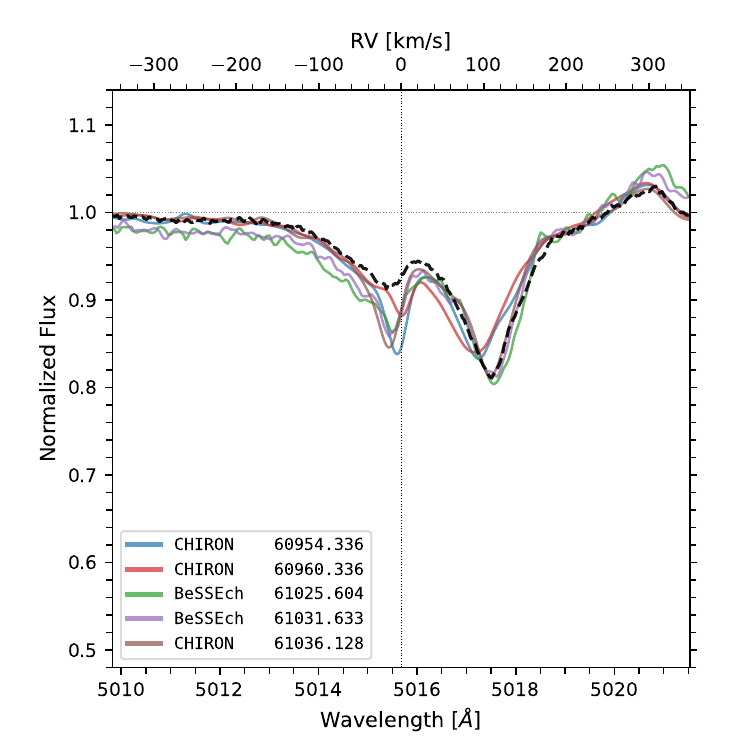}
\end{minipage}

\caption{
Transient shell absorption in \spec{He}{i}{5015} (neighboring \spec{Fe}{ii}{5018}) in two distinct states of the $V/R$ cycle in \Ha: $V>R$ with red-shifted shell absorption (left), and $V<R$ with blue-shifted shell absorption (right). The individual epochs (colored) are overplotted with an average of spectra not showing the transient shell absorption (dashed black; UVES on the left and CHIRON on the right). Even though the $V/R$ phase is similar between PUCHEROS and UVES spectra on the left, note also the large difference in the depth of the main shell absorption in \spec{Fe}{ii}{5018}, suggesting that the disk was overall denser during the PUCHEROS observations.
}
\label{fig:transient_shell_overplot}
\end{figure*}

Fig.~\ref{fig:transient_shell_overplot} shows \spec{He}{i}{5015} for all the spectra taken during transient shell phases, compared to the average line profile from the remaining spectra.

\end{appendix}

\end{document}